\documentclass[12pt, draftclsnofoot, onecolumn]{IEEEtran}
\usepackage[T1]{fontenc}
\usepackage[latin9]{inputenc}
\usepackage{color}
\usepackage{float}
\usepackage{textcomp}
\usepackage{bm}
\usepackage{amsmath}
\usepackage{amsthm}
\usepackage{amssymb}
\usepackage{graphicx}
\usepackage[unicode=true,
 bookmarks=true,bookmarksnumbered=true,bookmarksopen=true,bookmarksopenlevel=1,
 breaklinks=false,pdfborder={0 0 0},pdfborderstyle={},backref=false,colorlinks=false]
 {hyperref}
\hypersetup{pdftitle={Your Title},
 pdfauthor={Your Name},
 pdfpagelayout=OneColumn, pdfnewwindow=true, pdfstartview=XYZ, plainpages=false}

\makeatletter

\providecommand{\tabularnewline}{\\}
\floatstyle{ruled}
\newfloat{algorithm}{tbp}{loa}
\providecommand{\algorithmname}{Algorithm}
\floatname{algorithm}{\protect\algorithmname}

\theoremstyle{plain}
\newtheorem{thm}{\protect\theoremname}
\theoremstyle{remark}
\newtheorem{rem}[thm]{\protect\remarkname}

\usepackage[caption=false,font=footnotesize]{subfig}
\usepackage{bm}

\usepackage{cite}

\usepackage[T1]{fontenc}
\usepackage{algorithm}
\usepackage{algorithmic}
\IEEEoverridecommandlockouts

\@ifundefined{showcaptionsetup}{}{%
 \PassOptionsToPackage{caption=false}{subfig}}
\usepackage{subfig}
\makeatother

\providecommand{\remarkname}{Remark}
\providecommand{\theoremname}{Theorem}

\begin{document}
\title{Cooperative Data Collection with Multiple UAVs for Information Freshness
in the Internet of Things }
\author{Xijun~Wang, Mengjie~Yi, Juan~Liu, Yan~Zhang, Meng~Wang, and
Bo~Bai \vspace{-1.6cm}\thanks{Part of this work was presented at the IEEE INFOCOM WKSHPS, July 2020
\cite{yi2020deep}. }\thanks{X.~Wang is with School of Electronics and Information Technology,
Sun Yat-sen University, Guangzhou, 510006, China (e-mail: wangxijun@mail.sysu.edu.cn).
M. Yi is with School of Cyber Engineering, Xidian University, Xi\textquoteright an
710071, China (e-mail: mjyi@stu.xidian.edu.cn). J. Liu is with School
of Electrical Engineering and Computer Science, Ningbo University,
Zhejiang, 315211, China (e-mail: eeliujuan@gmail.com). Y. Zhang is
with State Key Lab of Integrated Service Networks, Information Science
Institute, Xidian University, Xi\textquoteright an, Shaanxi, 710071,
China (e-mail: yanzhang@xidian.edu.cn). M. Wang and B. Bai are with
Theory Lab, Central Research Institute, 2012 Labs, Huawei Technology
Co., Ltd., Hong Kong Science Park, Hong Kong SAR (e-mail: wang.meng7@huawei.com;
ee.bobbai@gmail.com).}\thanks{@ 2023 IEEE. Personal use of this material is permitted. Permission
from IEEE must be obtained for all other uses, including reprinting/republishing
this material for advertising or promotional purposes, collecting
new collected works for resale or redistribution to servers or lists,
or reuse of any copyrighted component of this work in other works.}}
\maketitle
\begin{abstract}
 Maintaining the freshness of information in the Internet of Things
(IoT) is a critical yet challenging problem. In this paper, we study
cooperative data collection using multiple Unmanned Aerial Vehicles
(UAVs) with the objective of minimizing the total average Age of Information
(AoI). We consider various constraints of the UAVs, including kinematic,
energy, trajectory, and collision avoidance, in order to optimize
the data collection process.  Specifically, each UAV, which has  limited
on-board energy, takes off from its initial location and flies over
sensor nodes to collect update packets in cooperation with the other
UAVs. The UAVs must land at their final destinations with non-negative
residual energy after the specified time duration to ensure they have
enough energy to complete their missions. It is crucial to design
the trajectories of the UAVs and the transmission scheduling of the
sensor nodes to enhance information freshness. We model the multi-UAV
data collection problem as a Decentralized Partially Observable Markov
Decision Process (Dec-POMDP), as each UAV is unaware of the dynamics
of the environment and can only observe a part of the sensors. To
address the challenges of this problem, we propose a multi-agent \textcolor{black}{Deep
Reinforcement Learning (DRL)-}based algorithm with centralized learning
and decentralized execution. In addition to the reward shaping, we
use  action masks to filter out  invalid actions and ensure that the
constraints are met. \textcolor{black}{Simulation results  demonstrate
that the proposed algorithms can significantly reduce the total average
AoI compared to the baseline algorithms, and the use of the action
mask method can improve the convergence speed of the proposed algorithm.}
\end{abstract}


\section{Introduction}

There is a rising need to support ubiquitous connections for Internet
of Things (IoT)-based applications, such as monitoring agricultural
growth, tracking marine life, and surveilling border areas, to name
but a few, in remote geographical areas where terrestrial network
infrastructures are limited. To meet this need, Unmanned Aerial
Vehicles (UAVs) have emerged as a promising technology for providing
timely, flexible, and elastic services in underserved areas due to
their high mobility and agility \cite{Mozaffari_TutorialUAV_WN_2019}\cite{zongyong_UAV5G_tutorial2019}.
Specifically, UAVs equipped with microprocessors and wireless transceivers
can dynamically move towards each sensor in IoT networks to collect
or disseminate information. With their fully controllable mobility
and high altitude, UAVs can exploit the Line-of-Sight (LoS) wireless
communications on the air-to-ground channel to improve throughput
and reduce transmission energy consumption. Moreover, UAVs can be
dispatched on demand and their locations can be promptly adjusted
according to the dynamic communication environment, providing fast
and flexible reconfiguration.

Due to their attractive characteristics, UAV-enabled IoT networks
have gained considerable attention and aroused numerous research interests
in recent years. However, most of the existing works have focused
on optimizing traditional performance metrics, such as system throughput,
coverage, and delay \cite{JGong_TminUAV_DC_2018,Yzeng_EEUAV_trajectory_optimization_2017,CHLiu_EE_FairCC_DRL_2018,RDing_3DTra-Freq-EE-FairC-DRL_2020TWC,PM_EE-VNF-DRL_TGCN2022}.
These performance metrics fail to quantify the freshness of information
in the UAV-enabled IoT networks, which is critical for mission-critical
applications \cite{Abd_elmagid-AoI_role_in_IoT-2019,BOM_TheRoleofUAV-IoTNetworkinFutureWildfireDetection_2021IoT},
such as forest fire containment and extinguishment, and disaster relief
surveillance and rescue. In these applications, the accuracy of the
decisions made using the collected information from IoT devices depends
heavily on the freshness of the information. This has led to the use
of the Age of Information (AoI) as a measure of fresh information
from the receiver's perspective \cite{S.Kaul_Mini_AoI_VehicularNet,AM_AoI-VNF-CompoundAC-MADRL-JSAC2021,Guo:2022ub}.

Due to the limited onboard energy of the UAV, it is difficult, if
not impossible, to collect data from all the sensors in a large area
within a stringent time requirement using a single UAV. Therefore,
it is necessary to study cooperative data collection with multiple
UAVs. However, compared with the single UAV-enabled IoT networks,
there are several technical challenges that result from having multiple
UAVs. Firstly, since the UAVs would collide with each other, the additional
collision avoidance constraint has to be guaranteed when designing
the UAVs' trajectories. Secondly, due to the mutual interference among
the simultaneous status updating, the scheduling of the IoT device
has to be carefully designed as well to mitigate transmission failure.
Thirdly, since the dynamic of the environment is affected by the actions
of all UAVs, the environment faced by each UAV is non-stationary.
Lastly, the action space increases exponentially with the number of
UAVs, making it more difficult to find an optimal decision.  \textcolor{black}{Previous
studies in this area have relied on global state information for coordination
\cite{JHu_UAVs_traj_DRL_2020,WFY_UAV-device-multiDRL_TCOM_2021},
which can lead to significant communication overhead. Therefore, further
study is needed on how to cooperatively collect status update packets
from sensors with multiple UAVs in a decentralized manner in order
to maintain information freshness.}

Motivated by these facts, we consider the IoT networks without terrestrial
network infrastructures and investigate the problem of cooperative
data collection with multiple UAVs for information freshness. In this
model, each UAV acts as an individual agent. Specifically, each UAV
with a limited amount of onboard energy is dispatched from its depot,
flies towards the sensors to collect status update packets, and finally
arrives at the destination within a fixed period. These UAVs must
avoid collisions during the flight and reach their destinations before
running out of energy. Each sensor node is equipped with a battery
and can harvest energy from the ambient environment. The sensor is
only available when it has enough energy. We jointly design the trajectories
of UAVs and the scheduling of sensors to minimize the total average
AoI.  The main contributions of this paper are summarized as follows.
\begin{itemize}
\item We formulate the multi-UAV data collection problem as a finite-horizon
Decentralized Partially Observable Markov Decision Process (Dec-POMDP)
due to the lack of terrestrial network infrastructures, the limited
observation of each UAV, and the unknown dynamics of the environment,
including the energy harvesting rate of the sensors and the line-of-sight
probability of the air-to-ground channel. 
\item We propose a multi-agent deep reinforcement learning (DRL) based algorithm
with Centralized Training with Decentralized Execution (CTDE) to cope
with the non-stationary environment. Particularly, the UAVs are jointly
trained by leveraging global information in the training phase, whereas
each UAV makes the decision independently based on its own observation
in the execution phase. Moreover, to ensure the kinematic, energy,
and trajectory constraints, we use an action mask to filter out the
invalid actions during the training and execution phases.
\item We conduct extensive simulations to evaluate the performance of the
proposed algorithm. The results show that the proposed algorithm can
effectively coordinate the trajectories of UAVs and the scheduling
of sensors. Compared with the baseline algorithms, the total average
AoI of the proposed algorithm is significantly reduced.
\end{itemize}

The paper is organized as follows: The related work is summarized
in Section \ref{sec:Related-Work}. The system model and problem formulation
are described in Section \ref{sec:System-Model}. The multi-agent
DRL approach is proposed in Section \ref{sec:Multi-agent-DRL}. The
simulation results and discussions are given in Section \ref{sec:Simulation-Results}.
Finally, we conclude this paper in Section \ref{sec:Conclusions}.

\section{Related Work\label{sec:Related-Work}}

\subsection{Single-UAV Data Collection}

The collection of fresh data in IoT networks using a single unmanned
aerial vehicle (UAV) has been extensively studied. In \cite{liMinimizingPacketExpiration2019},
the UAV's flight trajectory was designed to minimize the number of
expired packets based on Q-learning, taking into account the AoI deadline
constraint imposed on each sensor. In \cite{JLiu_UAV_AoIWSN}, two
age-optimal data collection problems were formulated to minimize either
the  average AoI or peak AoI, where the sensors are grouped into non-overlapping
clusters and the UAV flies along with the collection points to collect
data from a set of sensors. Both the charging time and trajectory
of the UAV were designed to minimize the average AoI for UAV-assisted
wirelessly powered IoT networks in \cite{Hu_UAV_AoI_chargeSN_IoT_2021},
where the UAV charges the sensors before collecting state updates
from them. In \cite{Jia_AoI_UAV_IoT_2019}, considering three data
acquisition strategies of UAV, i.e., hovering, flying, and hybrid,
a dynamic programming algorithm was used to optimize the visiting
order of the SNs and data acquisition strategies to minimize the average
AoI while also satisfying the energy constraints of the SN. The above
studies \cite{liMinimizingPacketExpiration2019,JLiu_UAV_AoIWSN,Hu_UAV_AoI_chargeSN_IoT_2021,Jia_AoI_UAV_IoT_2019}
all involve the UAV collecting status update packets from each sensor
only once, whereas in \cite{abd-elmagidDeepReinforcementLearning2019,Ferdowsi_tuyouhua-DRL-UAV-AoI_2021},
the UAV is allowed to visit each sensor multiple times within a given
time period. A DRL algorithm was proposed to optimize the UAV\textquoteright s
trajectory and the scheduling of sensors with the objective of minimizing
the normalized weighted sum of AoI in \cite{abd-elmagidDeepReinforcementLearning2019},
and this was extended in \cite{Ferdowsi_tuyouhua-DRL-UAV-AoI_2021}
by combining DRL and convex optimization.

However, the energy consumption of the UAV has been ignored in these
previous works. In reality, it is important to consider the UAV's
energy consumption when designing its flight trajectory due to the
limited onboard battery of the UAV. In our earlier work \cite{yi2020deep},
the age-optimal flight trajectory and transmission scheduling were
jointly designed considering both the energy and time constraints.
In \cite{abd-elmagidAveragePeakAgeofInformation2019}, the UAV was
used as a mobile relay and the average PAoI was minimized by jointly
optimizing the UAV\textquoteright s flight trajectory, energy allocations,
and transmission time durations at both the sensor and the UAV. In
\cite{Zhang_AoI_UAVrelay_energy_2020}, the sensing and communication
trade-off was studied in terms of time and energy consumption for
the cellularly connected UAV. In \cite{SMY_AoIEnergyAwareUAVAssistedDataCollectIoTNet:DRL_IoT2021},
the UAV's flight speed, hovering locations, and bandwidth allocation
were jointly optimized to minimize the weighted sum of the expected
average AoI, propulsion energy of the UAV, and the transmission energy
at IoT devices. 

\subsection{Multi-UAV Data Collection}

There have been several studies that involve the use of multiple UAVs
for collaborative data collection. In one study \cite{Samir_AoI_mltiUAV_transport_DDPG_2020},
UAVs were used to collect data generated by vehicles in intelligent
transportation systems and the Deep Deterministic Policy Gradient
(DDPG) algorithm was used to optimize the UAVs' trajectories and the
transmission scheduling of vehicles in order to minimize the expected
weighted sum AoI. Another study \cite{Abedin_multiUAV_maxEF_AoIconstraint_DQN_2020}
utilized the Deep Q-Network (DQN) to design the cooperative trajectories
of the UAVs in order to maximize total energy efficiency under both
energy and AoI constraints. While these studies both involve multiple
UAVs, all of the UAVs' actions are controlled by a single agent, which
can result in a very large action space. 

In contrast, the study \cite{JHu_UAVs_traj_DRL_2020} proposed a distributed
sense-and-send protocol in which each UAV acts as its own agent and
makes its own decisions based on a compound action actor-critic algorithm
in order to minimize the AoI. This approach was further extended in
\cite{WFY_UAV-device-multiDRL_TCOM_2021}, where a multi-UAV trajectory
design algorithm based on DDPG was proposed to minimize the AoI for
UAV-to-device communications. However, in these studies, each UAV
must communicate with the BS in order to obtain the full knowledge
of all other UAVs' states, leading to significant signaling overhead.

In \cite{OOS_MultiUAVAoI-WPCN-MultiagentRL_2021INFOCOMWKSH}, two
UAVs were used, one as an energy transmitter and the other as a data
collector, and the Independent DQN (IDQN) algorithm was employed to
design the UAVs' trajectories in order to minimize the AoI, enhance
energy transfer to devices, and minimize UAV energy consumption. With
IDQN, each UAV makes decisions based on its own observations, but
the convergence of IDQN is not guaranteed in a non-stationary multi-agent
environment. To address this issue, we model the collaborative data
collection problem as a Dec-POMDP and adopt the CTDE framework in
this work. Specifically, multiple UAVs are centrally trained with
global state information to ensure convergence, and then each UAV
makes its decisions based on local observations without the need for
a central entity to collect and disseminate global state information.

\section{System Model and Problem Formulation\label{sec:System-Model}}

In this section, we first describe the multi-UAV enabled IoT network.
Then, we present the energy consumption model, the air-to-ground channel
model, and the evolution of the age of information. Finally, we formulate
a data collection problem by optimizing the flight trajectories of
the UAVs and the transmission scheduling of the sensors.

\subsection{Scenario Description}

\begin{figure}[t]
\centering

\includegraphics[width=0.6\textwidth]{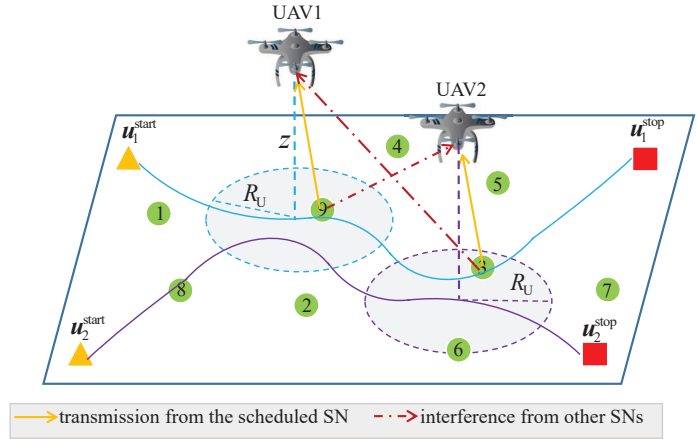}\caption{\label{fig:system-model}An illustration of multi-UAV enabled data
collection.}
\end{figure}

As shown in Fig. \ref{fig:system-model}, we consider a multi-UAV
enabled IoT network containing $N$ randomly distributed sensor nodes
(SNs).   The set of all  SNs is denoted by $\mathcal{N}=\{1,2,\ldots,N\}$
and the location of SN $n\in\mathcal{N}$ is represented by $\bm{w}_{n}=(x_{n},y_{n},0)$
in a three-dimensional Cartesian coordinate system. A swarm of $M$
rotary-wing UAVs is employed to cooperatively collect status updates
from the SNs, which are directly used by the UAVs to execute certain
tasks. The set of all  UAVs is denoted as $\mathcal{M}=\{1,2,\ldots,M\}$.
 We assume a discrete-time system where time is divided into equal-length
time slots, each lasting  $\tau_{0}$ seconds. The UAVs  are required
to support collaborative data collection for a total of $T$ time
slots.\footnote{The value of $T$ depends on the specific task of the UAVs.}
In particular, UAV $m\,(m\in\mathcal{M})$ takes off from an initial
location $\boldsymbol{u}_{m}^{\text{start}}$ and flies over various
SNs to collect their status updates. By the end of the $T$-th slot,
the UAV needs to arrive at its final destination $\boldsymbol{u}_{m}^{\text{stop}}$.

\textcolor{black}{At the beginning of time slot $t$, the location
of UAV $m\in\mathcal{M}$ in the three-dimensional Cartesian coordinate
is denoted by $\boldsymbol{u}_{m}(t)=(x_{m}(t),y_{m}(t),z_{m}(t))$,
where $(x_{m}(t),y_{m}(t))$ is the projection of the location of
UAV $m$ on the ground and $z_{m}(t)$ is the altitude of UAV $m$.
We assume that all UAVs fly at the same fixed altitude, i.e., $z_{m}(t)=z,\forall t$.
The velocity of UAV $m$ at the beginning of time slot $t$ is represented
in polar coordinates and denoted by $\boldsymbol{v}_{m}(t)=(v_{m}^{s}(t),\varphi_{m}(t))$,
where $v_{m}^{s}(t)=\left\Vert \boldsymbol{v}_{m}(t)\right\Vert $
is the speed of UAV $m$ and $\varphi_{m}(t)$ is the velocity direction
with $0\leq\varphi_{m}(t)\leq2\pi$. We assume a constant acceleration
$\bm{a}_{c,m}(t)=\dot{\bm{v}}_{m}(t)$ during one time slot and hence
the velocity can be updated by $\bm{v}_{m}(t+1)=\bm{v}_{m}(t)+\bm{a}_{c,m}(t)\tau_{0}$.
As the quad-rotor UAV is able to easily steer by adjusting the rotation
rate of four rotors, we assume that the UAV can change its direction
instantly at the beginning of a time slot, and the flight direction
is then fixed for the rest of the time slot. In practice, UAVs are
subject to kinematic constraints. In particular, the speed of the
UAV $m$ at the slot $t$ cannot exceed its maximum value $v_{\max}^{s}$,
i.e., $v_{m}^{s}(t)\leq v_{\max}^{s}$, and the turning angle of UAV
$m$ at the slot $t$, $\triangle\varphi_{m}(t)=\varphi_{m}(t)-\varphi_{m}(t-1)$,
cannot exceed its maximum value $\triangle\varphi_{\max}$, i.e.,
$\mid\triangle\varphi_{m}(t)\mid\leq\triangle\varphi_{\max}$.}\footnote{\textcolor{black}{This work can also be extended to the 3D mobility
of the UAVs. In particular, the velocity of the UAV is denoted by
$\boldsymbol{v}_{m}(t)=(v_{m}^{s}(t),\varphi_{p,m}(t),\varphi_{a,m}(t))$,
where $v_{m}^{s}(t)=\left\Vert \bm{v}_{m}(t)\right\Vert $ is the
speed of UAV $m$ and $(\varphi_{p,m}(t),\varphi_{a,m}(t))$ is the
direction of the UAV.  $\varphi_{p,m}(t)$ is the polar angle from
the positive $z$-axis with $0\leq\varphi_{p,m}(t)\leq\pi$, and $\varphi_{a,m}(t)$
is the azimuthal angle in the $xy$-plane from the positive x-axis
with $-\pi\leq\varphi_{a,m}(t)\leq\pi$. The kinematic constraints
on the speed and the turning angle are similar to those in 2D mobility.
In addition, the altitude of the UAV is limited between $z_{\textrm{min}}$
and $z_{\textrm{max}}$.}}\textcolor{black}{{}  We denote the velocities of all UAVs as $\boldsymbol{V}=(\bm{v}_{1},\bm{v}_{2},\ldots,\bm{v}_{M})$,
where $\bm{v}_{m}=(\boldsymbol{v}_{m}(1),\boldsymbol{v}_{m}(2),\ldots,\boldsymbol{v}_{m}(T))$
represents the sequence of velocities of UAV $m$. The flight trajectory
of UAV $m$ is defined as a sequence of locations it flies over, i.e.,
$\bm{p}_{m}=(\bm{u}_{m}(1),\bm{u}_{m}(2),\ldots,\boldsymbol{u}_{m}(T))$,
where $\bm{u}_{m}(1)=\bm{u}_{m}^{\text{start}}$ and $\bm{u}_{m}(T)=\bm{u}_{m}^{\textrm{stop}}$.
} To avoid collisions between the UAVs, the distance between any
two UAVs at time slot $t$ cannot be less than the safe distance,
i.e., $d_{m,m'}(t)\geq d_{\textrm{safe}}$, for any $m\neq m'(m,m'\in\mathcal{M})$. 

Each UAV has a coverage area on the ground with radius $R_{U},m\in\mathcal{M}$,
which is fixed and determined by the transmit power of the SN, the
altitude of the UAV, the antenna gains, and the channel path loss.
In particular, we set the radius as $R_{U}=\sqrt{d^{2}-z^{2}},$ where
$d=\frac{c}{4\pi f_{c}}\left(\frac{P_{c}G_{s}G_{u}}{\xi_{\text{th}}\sigma^{2}\eta_{\text{NLoS}}}\right)^{1/\varsigma}$.\textcolor{black}{}\footnote{In this study, we consider the coverage area as a circle with a fixed
radius $R_{U}$, which is determined by the NLoS path loss path-loss
coefficient $\eta_{NLoS}$, for simplicity. This is because NLoS conditions
typically result in higher path loss, and any SU with a LoS channel
within this circle would also be covered by the UAV. However, we acknowledge
that considering both LoS and NLoS path loss is crucial for a more
accurate representation of the coverage area. While our approach is
useful for a basic evaluation, it can be further extended to a more
general model with a varying coverage area.}\textcolor{black}{{} The specific notations will be explained in Subsection
\ref{subsec:Air-to-Ground-Channel}. }Let $\overline{d}_{n,m}(t)$
denote the distance between SN $n$ and the projection of UAV $m$
on the ground at the beginning of slot $t$. If $\overline{d}_{n,m}(t)\leq R_{U}$,
then SN $n$ is within the coverage area of UAV $m$ at slot $t$.
At every time slot, each UAV has to decide which SN within its coverage
area is scheduled to update its status. Let $\boldsymbol{b}_{m}=(b_{m}(1),b_{m}(2),\ldots,b_{m}(T))$
be the SN scheduling vector of UAV $m$, where $b_{m}(t)=n$ indicates
that SN $n$ transmits a status update packet to UAV $m$ at time
slot $t$ and $b_{m}(t)=0$ means that no SN is scheduled by UAV $m$
at time slot $t$. The SN scheduling vector of all UAVs is represented
by $\boldsymbol{B}=(\boldsymbol{b}_{1},\boldsymbol{b}_{2},\ldots,\boldsymbol{b}_{M})$.

\subsection{Energy Model}

\subsubsection{UAV Energy Model}

We assume that all the UAVs have the same amount of initial onboard
energy, denoted by $E_{\text{max}}$. The major energy consumption
of the UAV consists of communication energy and propulsion energy.
Since each UAV acts as a receiver for the status update packets, its
communication energy consumption is much smaller than the propulsion
energy consumption and hence can be ignored. \textcolor{black}{The
propulsion energy consumption of UAV $m$ $E_{m}(t)$ in time slot
$t$ can be expressed as follows \cite{RDing_3DTra-Freq-EE-FairC-DRL_2020TWC}:
\begin{align}
E_{m}(t)= & \tau_{0}n_{r}\Bigg[\frac{\sigma}{8}\left(\frac{T_{m}^{\textrm{h}}(t)}{c_{T}\rho A}+3\left(v_{m}^{s}(t)\right)^{2}\right)\sqrt{\frac{T_{m}^{\textrm{h}}(t)\rho c_{s}^{2}A}{c_{T}}}+\frac{1}{2}d_{0}\rho c_{s}A\left(v_{m}^{s}(t)\right)^{3}+\nonumber \\
 & \qquad(1+c_{f})T_{m}^{\textrm{h}}(t)\left(\sqrt{\frac{\left(T_{m}^{\textrm{h}}(t)\right)^{2}}{4\rho^{2}A^{2}}+\frac{\left(v_{m}^{s}(t)\right)^{4}}{4}}-\frac{\left(v_{m}^{s}(t)\right)^{2}}{2}\right)^{\frac{1}{2}}\Bigg],\label{eq:Energy_Consumption}
\end{align}
where $n_{r}$ is the number of rotors, $\sigma$ is the local blade
section drag coefficient, $c_{T}$ indicates the thrust coefficient
based on disk area, $\rho$ represents the density of air, $A$ and
$c_{s}$ denote the disk area for each rotor and rotor solidity, respectively,
$d_{0}$ is the fuselage drag ratio for each rotor,  $c_{f}$ is the
incremental correction factor of induced power, and $T_{m}^{\textrm{h}}(t)$
is the thrust of each rotor. For better exposition, we consider only
the acceleration in a straight line with the velocity and omit the
acceleration component perpendicular to the velocity \cite{RDing_3DTra-Freq-EE-FairC-DRL_2020TWC}.
Then, the thrust of each rotor can be expressed as
\begin{equation}
T_{m}^{\textrm{h}}(t)=\frac{1}{n_{r}}\left[\left(Wa_{c,m}(t)+\frac{1}{2}\rho\left(v_{m}^{s}(t)\right)^{2}S_{FP}\right)^{2}+(Wg)^{2}\right]^{1/2},\label{eq:Thrust_rotor}
\end{equation}
where $a_{c,m}(t)=(v_{m}^{s}(t+1)-v_{m}^{s}(t))/\tau_{0}$, $S_{FP}$
represents the fuselage equivalent flat plate area, $W$ denotes the
weight of the UAV, and $g$ is the gravity acceleration.}

\subsubsection{SN Energy Model}

We consider that each SN has a battery with finite capacity $E_{\textrm{max}}^{\textrm{sn}}$
and is able to harvest energy from the ambient environment. Let $E_{n}^{\textrm{sn}}(t)$
denote the battery level of SN $n$ at the beginning of slot $t$.
Without loss of generality, we assume that the batteries of all SNs
are fully charged at the start, i.e., $E_{n}^{\textrm{sn}}(1)=E_{\textrm{max}}^{\textrm{sn}}$
for any $n\in\mathcal{N}$. \textcolor{black}{We also assume that
SNs are able to continuously harvest energy from sources such as solar
and wind \cite{sn-energy-bonuli1_JSAC2016,sn-energy-bonuli2_JSAC2016,sn-energy-bonuli3_TIT2020}.
 However, since the ambient environment's energy is weather-dependent
and unreliable, we model the harvested energy at each SN as  an independent
 Bernoulli process with parameter $\lambda_{n}$. This means that
at each time slot, there is a probability $\lambda_{n}$ that a certain
amount of energy will arrive at SN $n$. Moreover, since the renewable
energy generators (e.g., solar panels) used by SNs to harvest energy
are independent of the transceivers of SNs, the SNs can simultaneously
harvest energy and transmit information.} We denote this arrival
of energy with an indicator  $\kappa_{n}(t)$,  where $\kappa_{n}(t)=1$
indicates that the amount of $E_{n}^{\textrm{har}}$ energy is harvested
by SN $n$ at slot $t$ and $\kappa_{n}(t)=0$ otherwise. We assume
that the scheduled SN transmits a status update packet in a time slot
with a fixed power $P_{c}$, and hence the energy required by SN $n$
for status updating is $E_{n}^{\textrm{c}}=P_{c}\tau_{0}$. Let $\zeta_{n}(t)$
be an indicator that denotes whether SN $n$ is scheduled to transmit
at slot $t$, where $\zeta_{n}(t)=1$ if it is scheduled by any UAV,
and $\zeta_{n}(t)=0$ otherwise. Specifically, we have $\zeta_{n}(t)=1-\prod_{m=1}^{M}\bm{1}(b_{m}(t)\neq n)$.
It is worth noting that the SN can be scheduled only if it has enough
energy in the battery, i.e., $E_{n}^{\textrm{sn}}(t)\geq E_{n}^{c}$.
Therefore, the dynamics of the battery level at SN $n$ are given
by 
\begin{equation}
E_{n}^{\textrm{sn}}(t+1)=\begin{cases}
\min\{E_{n}^{\textrm{sn}}(t)+\kappa_{n}(t)E_{n}^{\textrm{har}}-E_{n}^{\textrm{c}},E_{\textrm{max}}^{\textrm{sn}}\}, & \text{if }\zeta_{n}(t)=1,\\
\min\{E_{n}^{\textrm{sn}}(t)+\kappa_{n}(t)E_{n}^{\textrm{har}},E_{\textrm{max}}^{\textrm{sn}}\}, & \textrm{otherwise}.
\end{cases}\label{eq:energySN_update}
\end{equation}

\subsection{Air-to-Ground Channel\label{subsec:Air-to-Ground-Channel}}

The obstacles in the environment can impact the air-to-ground (A2G)
signal propagation. Depending on the specific propagation environment,
the A2G channel can be either LoS or non-line-of-sight (NLoS). However,
the information about the exact locations, heights, and number of
obstacles is generally not available in practical scenarios. The UAV
will have a LoS view towards a specific SN with a given probability.
\textcolor{blue}{{} }Let $d_{n,m}(t)$ denote the Euclidean distance
between SN $n$ and UAV $m$ in slot $t$. Then, the LoS probability
between SN $n$ and UAV $m$ is given by \cite{al2014optimal}
\begin{equation}
p_{n,m}^{\text{LoS}}(t)=\frac{1}{1+\beta_{0}\exp\left(\text{\textminus}\beta_{1}\left(\frac{180}{\pi}\arcsin\left(\frac{z}{d_{n,m}(t)}\right)\text{\textminus}\beta_{0}\right)\right)},
\end{equation}
where $\beta_{0}$ and $\beta_{1}$ are constants determined by the
environment. Then, the path loss between SN $n$ and UAV $m$ in slot
$t$ can be expressed as
\begin{equation}
PL_{n,m}(t)=\begin{cases}
\left(\frac{4\pi f_{c}d_{n,m}(t)}{c}\right)^{\varsigma}\eta_{\text{LoS}}, & \textrm{w.p. \ensuremath{p_{n,m}^{\text{LoS}}(t)}},\\
\left(\frac{4\pi f_{c}d_{n,m}(t)}{c}\right)^{\varsigma}\eta_{\text{NLoS}}, & \textrm{otherwise.}
\end{cases}\label{eq:path-loss}
\end{equation}
where $\varsigma$ is the path loss exponent, $f_{c}$ is the carrier
frequency, $c$ represents the speed of light in vacuum, and $\eta_{\text{LoS}}$
and $\eta_{\text{NLoS}}$ ($\eta_{\text{NLoS}}>\eta_{\text{LoS}}>1$)
are the excessive path-loss coefficients for LoS and NLoS links, respectively.
We assume that the UAV does not know whether the channel state is
LoS or NLoS before the scheduling, and is unaware of the LoS probabilities
either.

Each status update packet is assumed to be transmitted in a single
time slot.\textcolor{black}{{} As in \cite{Zhangrui_slot_dis_unchange_TWC2018},
we choose the slot length $\tau_{0}$ in such a way that the distance
between the UAV and the SNs can be assumed to be approximately constant
within each slot. For instance, $\tau_{0}$ might be chosen such that
$\tau_{0}v_{\text{max}}^{s}\ll z$. }Since there are multiple UAVs
collecting status updates from the SNs in the same frequency band,
the  SNs transmitting in the same time slot may interfere with each
other. \textcolor{black}{We assume that both the SNs and UAVs are
equipped with  omni-directional antennas, with antenna gains  denoted
as $G_{s}$ and $G_{u}$, respectively. Without loss of generality,
we set $G_{s}=G_{u}=0\thinspace$dB. It is noteworthy that this work
can be extended to  scenarios with directional antennas.}\textcolor{blue}{}
Then, the signal-to-interference-noise ratio (SINR) of the A2G channel
between SN $n$ and UAV $m$ is given by
\begin{align}
\xi_{n,m}(t) & =\frac{P_{c}G_{s}G_{u}\gamma_{n,m}(t)}{\sigma^{2}+I_{n,m}(t)},\label{eq:SINR}
\end{align}
where $P_{c}$ is the transmission power of each SN, $\gamma_{n,m}(t)=\frac{1}{PL_{n,m}(t)}$
is the channel gain, and $\sigma^{2}$ is the noise power. In (\ref{eq:SINR}),
$I_{n,m}(t)=\sum_{n'\in\mathcal{N}_{-m}(t)/n}P_{c}\gamma_{n',m}(t)$
is the total interference power, where $\mathcal{N}_{-m}(t)$ is the
set of all SNs that are scheduled by other UAVs $m'$ (i.e., $m'\in\mathcal{M},m'\neq m$)
in slot $t$. Then, the packet from SN $n$ is successfully received
by UAV $m$ in slot $t$ if the SINR $\xi_{n,m}(t)$ is no smaller
than a threshold $\xi_{\text{th}}$; otherwise, it fails.

\subsection{Age of Information}

We consider that the UAVs directly utilize the collected status updates
to make subsequent decisions as in \cite{Ferdowsi_tuyouhua-DRL-UAV-AoI_2021}
and employ AoI to measure the freshness of information at the UAVs.
In particular, the AoI of SN $n$ $\delta_{n}(t)$ is defined as the
time elapsed since the generation of the latest status update received
by any UAV.  We consider a generate-at-will policy, where an SN
generates a status update packet whenever it is scheduled. \textcolor{black}{The
status update packets contain information about the status of the
physical process of interest and the time instant (i.e., the time
when the sample was generated).} If UAV $m$ schedules SN $n$ to
update its status at time slot $t$ and the transmission is successful
(i.e., $\xi_{n,m}(t)\geq\xi_{\text{th}}$), the AoI of SN $n$ is
decreased to one (due to the one time slot used for packet transmission);
otherwise, it is increased by one. Then, the dynamic of the AoI of
SN $n$ is expressed as 
\begin{equation}
\delta_{n}(t+1)=\begin{cases}
1, & \text{if }\zeta_{n}(t)=1\textrm{ and \ensuremath{\xi_{n,m}(t)\geq\xi_{\text{th}},}}\\
\textrm{min}(\delta_{n}(t)+1,\delta_{\max}), & \text{otherwise},
\end{cases}\label{eq:update-AOI}
\end{equation}
where $\delta_{\text{max}}$ is the maximum value of AoI. \textcolor{black}{The
reason we set an upper limit of the AoI is that the highly outdated
status update will not be of any use to time-critical IoT applications
\cite{AoI-max_IoT2022,Abd_elmagid-AoI_role_in_IoT-2019}. Moreover,
the value of $\delta_{\max}$ is application-dependent and can be
small or large.}\textcolor{blue}{{} }We also assume that the AoI of
each SN is maintained by the SN itself, as a successful transmission
is acknowledged immediately by the UAV.

\subsection{Problem Formulation}

In this multi-UAV enabled data collection problem, our objective is
to minimize the time-average total expected AoI, which is the time-average
of the sum of the expected AoI associated with each SN, by jointly
optimizing the trajectories of UAVs and the scheduling of SNs. The
optimization problem can be expressed as follows:\begin{subequations}
\begin{align}
\text{P1: }\min_{\boldsymbol{V},\boldsymbol{B}}\quad & \frac{1}{T}\mathbb{E}\left[\sum_{t=1}^{T}\sum_{n=1}^{N}\delta_{n}(t)\right],\label{eq:object function}\\
\text{s.t.}\quad & d_{m,m'}(t)\geq d_{\textrm{safe}},\forall m,m'\in\mathcal{M},m\neq m',\label{eq:collision}\\
 & 0\leq v_{m}^{s}(t)\leq v_{\max}^{s},\forall m\in\mathcal{M},\label{eq:speed_constra}\\
 & 0\leq\varphi_{m}(t)\leq2\pi,\forall m\in\mathcal{M},\label{eq:angle_constra}\\
 & |\triangle\varphi_{m}(t)|\leq\triangle\varphi_{\max},\forall m\in\mathcal{M},t\geq2,\label{eq:turning_angle_constra}\\
 & \zeta_{n}(t)\leq\bm{1}(E_{n}(t)\geq E_{n}^{c},\overline{d}_{n,m}(t)\leq R_{\textrm{U}}),\forall n\in\mathcal{N},\exists m\in\mathcal{M},\label{eq:SN_schedule}\\
 & \sum_{t=1}^{T}E_{m}(t)\leq E_{\text{max}},\forall m\in\mathcal{M},\label{eq:energy}\\
 & \bm{u}_{m}(1)=\bm{u}_{m}^{\textrm{start}},\bm{u}_{m}(T)=\bm{u}_{m}^{\textrm{stop}},\forall m\in\mathcal{M}.\label{eq:start_stop_ps}
\end{align}
\end{subequations}The collision avoidance constraint is given in
(\ref{eq:collision}).\textcolor{black}{{} The speed constraint (\ref{eq:speed_constra})
and direction constraints (\ref{eq:angle_constra}) and (\ref{eq:turning_angle_constra})
ensure that the UAV satisfies the kinematic constraints.} The SN
scheduling constraint (\ref{eq:SN_schedule}) ensures that each SN
can be scheduled only if the remaining energy in the battery is enough
for transmission and if it is within the coverage area of at least
one UAV. The UAV energy constraint (\ref{eq:energy}) ensures that
each UAV will not run out of energy before the end of $T$-th slot.
The trajectory constraint (\ref{eq:start_stop_ps}) guarantees
that each UAV starts from its initial location at the first slot and
arrives at its final location at the end of $T$-th slot. Solving
the above stochastic optimization problem is very challenging due
to the unknown of the environmental dynamics, the causality of energy
consumption, and the limited observation capability of the UAVs.
 The traditional optimization methods that require the knowledge
of all environmental dynamics do not apply to this problem. Instead,
we propose a multi-agent DRL approach to jointly design the trajectory
planning and the transmission scheduling.

\section{Multi-agent DRL Approach\label{sec:Multi-agent-DRL}}

In this section, we consider the multi-UAV enabled data collection,
where a group of UAVs cooperatively collect status updates from the
SNs. Each UAV works as an independent agent, making its own decisions
without sharing information with other UAVs. Since each UAV does not
know the actions of other UAVs, it is only able to observe the AoI
and the battery level of the SNs in its vicinity. As such, we first
cast the multi-UAV enabled data collection problem into a Dec-POMDP,
which enables multiple separate POMDPs operating on various agents
to behave independently while working towards an objective function
that relies on the behaviors of all the agents \cite{oliehoek2016concise}.
Then, we propose a multi-agent DRL-based algorithm to obtain the policy
for each UAV.

\subsection{Dec-POMDP Formulation}

In the Dec-POMDP, there are multiple UAVs interacting with the environment
represented by a set of states $\mathcal{S}$. Each UAV is controlled
by its own dedicated agent, and the objective function depends on
the actions of all the agents. At each time slot, each UAV receives
its own observation and takes its individual action. A cost is received
accordingly by each UAV, and then the environment transits to a new
state. In the following, we define the state, observation, action,
state transition, and cost of the Dec-POMDP in more detail.

\subsubsection{State}

The state in slot $t$ is defined as $\boldsymbol{s}(t)=(\boldsymbol{u}(t),\boldsymbol{\delta}(t),\boldsymbol{V}(t),\boldsymbol{\varphi}(t-1),\boldsymbol{E}^{\textrm{sn}}(t),\boldsymbol{\phi}(t),\boldsymbol{\psi}(t))\in\mathcal{S}$,
which consists of seven parts:
\begin{itemize}
\item $\boldsymbol{u}(t)=(\boldsymbol{u}_{1}(t),\boldsymbol{u}_{2}(t),\ldots,\boldsymbol{u}_{M}(t))$
is the locations of all UAVs at the beginning of time slot $t$.
\item $\boldsymbol{\delta}(t)=(\delta_{1}(t),\delta_{2}(t),\ldots,\delta_{N}(t))$
is the AoI of all  SNs in time slot $t$. 
\item $\boldsymbol{V}(t)=(v_{1}^{s}(t),v_{2}^{s}(t),\ldots,v_{M}^{s}(t))$
is the speed of all UAVs at the beginning of time slot $t$.
\item $\boldsymbol{\varphi}(t-1)=(\varphi_{1}(t-1),\varphi_{2}(t-1),\ldots,\varphi_{M}(t-1))$
is the direction of velocity of all UAVs at time slot $t-1$.
\item $\boldsymbol{E}^{\textrm{sn}}(t)=(E_{1}^{\textrm{sn}}(t),E_{2}^{\textrm{sn}}(t),\ldots,E_{N}^{\textrm{sn}}(t))$
is the battery level of all SNs in time slot $t$.
\item \textcolor{black}{$\boldsymbol{\phi}(t)=(\phi_{1}(t),\phi_{2}(t),\ldots,\phi_{M}(t))$
is the time difference of all UAVs at the beginning of slot $t$.
Particularly, $\phi_{m}(t)=T-t+1-T_{m}^{\text{req}}(t)$ is the difference
between the remaining time of the flight cruise, i.e., $T-t+1$, and
the time required by UAV $m$ to reach the final destination, i.e.,
$T_{m}^{\text{req}}(t)$. We assume that when the speed of UAV $m$
is zero, its velocity direction can be any value between  (0, 2$\pi$).
There are two cases for calculating $T_{m}^{\text{req}}(t)$. In the
first case, if UAV $m$ can immediately turn its velocity direction
towards the destination, i.e., $\triangle\varphi_{m}^{\textrm{stop}}(t)\leq\triangle\varphi_{\max}$
or $v_{m}^{s}(t)=0$, then $T_{m}^{\text{req}}(t)$ consists of the
number of time slots required to accelerate to the maximum speed and
then fly to the destination with the maximum speed. In the other case,
if UAV $m$ cannot turn its velocity direction immediately towards
the destination, we stipulate that it first decreases its speed to
zero, then turns its velocity direction towards the destination and
accelerates to the maximum speed, and then flies to the destination
with the maximum speed. Therefore, $T_{m}^{\text{req}}(t)$ consists
of the number of time slots required to decelerate its speed to zero,
accelerate to the maximum speed, and then fly to the destination with
the maximum speed. For computational and presentation simplicity,
when the UAV $m$ cannot immediately turn the velocity direction towards
the destination, we assume that the velocity direction is in the opposite
direction to the final position. By doing this, we can obtain an upper
bound on $T_{m}^{\text{req}}(t)$. As a result, $T_{m}^{\text{req}}(t)$
can be expressed as
\begin{equation}
T_{m}^{\text{req}}(t)=\begin{cases}
1+\lceil\frac{\left\Vert \bm{u}_{m}(t)-\bm{u}_{m}^{\text{stop}}\right\Vert -\triangle_{m,1}(t)}{v_{\max}^{s}\tau_{0}}\rceil, & \textrm{if }\triangle\varphi_{m}^{\textrm{stop}}(t)\leq\triangle\varphi_{\max}\textrm{ or }v_{m}^{s}(t)=0,\\
2+\lceil\frac{\left\Vert \bm{u}_{m}(t)-\bm{u}_{m}^{\text{stop}}\right\Vert +\triangle_{m,2}(t)-\triangle_{m,3}}{v_{\max}^{s}\tau_{0}}\rceil, & \textrm{otherwise,}
\end{cases}\label{eq:-7}
\end{equation}
where $\lceil\cdotp\rceil$ denotes the ceiling, $\triangle\varphi_{m}^{\textrm{stop}}(t)=|\varphi_{m}(t)-\varphi_{m}^{\textrm{stop}}(t)|$
denotes the absolute value of the angle difference between the direction
of the velocity of UAV $m$ $\varphi_{m}(t)$ and the direction of
the vector of the current position of UAV $m$ pointing to the destination
$\varphi_{m}^{\textrm{stop}}(t)$. $\triangle_{m,1}(t)=\frac{v_{\max}^{s}+v_{m}^{s}(t)}{2}\tau_{0}$
indicates the distance flown by UAV $m$ from $v_{m}^{s}(t)$ to the
maximum speed $v_{\max}^{s}$, $\triangle_{m,2}(t)=\frac{v_{m}^{s}(t)}{2}\tau_{0}$
indicates the distance flown by UAV $m$ from $v_{m}^{s}(t)$ to zero,
and $\triangle_{m,3}=\frac{v_{\max}^{s}}{2}\tau_{0}$ denotes the
distance flown by UAV $m$ from zero velocity to $v_{\max}^{s}$.
}
\item $\boldsymbol{\psi}(t)=(\psi_{1}(t),\psi_{2}(t),\ldots,\psi_{M}(t))$
is the energy difference of all UAVs at the beginning of slot $t$.\textcolor{black}{{}
Particularly, $\psi_{m}(t)=E_{\text{max}}-\sum_{i=1}^{t-1}E_{m}(i)-E_{m}^{\text{req}}(t)$
is the difference between the remaining energy of UAV $m$, i.e.,
$E_{\text{max}}-\sum_{i=1}^{t-1}E_{m}(i)$, and the energy required
for the UAV $m$ to reach the destination from its current location,
i.e., 
\begin{equation}
E_{m}^{\textrm{req}}(t)=\begin{cases}
E_{m}^{a_{c,m}^{1}(t)}+\sum_{i=1}^{T_{m}^{\text{req}}(t)-1}E_{m}^{v_{\max}^{s},0}, & \textrm{if }\triangle\varphi_{m}^{\textrm{stop}}(t)\leq\triangle\varphi_{\max}\textrm{ or }v_{m}^{s}(t)=0,\\
E_{m}^{a_{c,m}^{2}(t)}+E_{m}^{0,a_{c,m}^{3}}+\sum_{i=1}^{T_{m}^{\text{req}}(t)-2}E_{m}^{v_{\max}^{s},0}, & \textrm{otherwise},
\end{cases}\label{eq:}
\end{equation}
where $E_{m}^{a_{c,m}^{1}(t)}$ denotes the energy consumed in slot
$t$ when $a_{c,m}(t)=\frac{v_{\max}^{s}-v_{m}^{s}(t)}{\tau_{0}}$,
$E_{m}^{a_{c,m}^{2}(t)}$ denotes the energy consumed in slot $t$
when $a_{c,m}(t)=\frac{-v_{m}^{s}(t)}{\tau_{0}}$, $E_{m}^{0,a_{c,m}^{3}}$
denotes the energy consumed in one time slot when $v_{m}^{s}(t)=0$
and $a_{c,m}(t)=\frac{v_{\max}^{s}}{\tau_{0}}$, and $E_{m}^{v_{\max}^{s},0}$
denotes the energy consumed in one time slot when $v_{m}^{s}(t)=v_{\max}^{s}$
and $a_{c,m}(t)=0.$  }
\end{itemize}

\subsubsection{Observation}

\textcolor{black}{We assume that the SNs will report their AoI and
battery level at the beginning of each time slot using orthogonal
access and each UAV can receive these information from the SNs locating
within its coverage area.}\textcolor{blue}{{} }The observation of UAV
$m$ at slot $t$ is represented as $\boldsymbol{o}_{m}(t)=(\bm{u}_{m}(t),\hat{\bm{\delta}}_{m}(t),$
$v_{m}^{s}(t),$ $\varphi_{m}(t-1),\hat{\bm{E}}_{m}^{\text{sn}}(t),\phi_{m}(t),\psi_{m}(t))$,
$\forall m\in\mathcal{M}$, where $\hat{\boldsymbol{\delta}}_{m}(t)=(\hat{\delta}_{m,1}(t),\hat{\delta}_{m,2}(t),\ldots,\hat{\delta}_{m,N}(t))$
and $\hat{\boldsymbol{E}}_{m}^{\text{sn}}(t)=(\hat{E}_{m,1}^{\text{sn}}(t),\hat{E}_{m,2}^{\text{sn}}(t),\ldots,\hat{E}_{m,N}^{\text{sn}}(t))$
denote, respectively, the AoI and the battery level of the SNs which
can be observed by UAV $m$ at time slot $t$. In particular, if SN
$n$ is in the coverage of UAV $m$ at time slot $t$, we have $\hat{\delta}_{m,n}(t)=\delta_{n}(t)$
and $\hat{E}_{m,n}^{\text{sn}}(t)=E_{n}^{\text{sn}}(t)$; otherwise,
$\hat{\delta}_{m,n}(t)=\textrm{none}$ and $\hat{E}_{m,n}^{\text{sn}}(t)=\text{none}$.

\subsubsection{Action}

\textcolor{black}{The action of UAV $m$ in time slot $t$, $\boldsymbol{a}_{m}(t)=(v_{m}^{s}(t+1),\varphi_{m}(t),b_{m}(t))$,
is characterized by its speed $v_{m}^{s}(t+1)$ at the beginning of
time slot $t+1$, the direction of velocity $\varphi_{m}(t)$  and
the scheduling of SNs $b_{m}(t)$ in time slot $t$. We discretize
the values of $v_{m}^{s}(t+1)$ and $\varphi_{m}(t)$ as  $v_{m}^{s}(t+1)\in\{0,\frac{1}{N_{1}}v_{\max}^{s},\frac{2}{N_{1}}v_{\max}^{s},\ldots,v_{\max}^{s}\}$
and $\varphi_{m}(t)\in\{0,\frac{1}{N_{2}}2\pi,\frac{2}{N_{2}}2\pi,\ldots,2\pi\}$,
respectively, where $N_{1}$ and $N_{2}$ are  positive integers.
Therefore, the action space of each UAV is defined as $\mathcal{A}=\mathcal{V}^{s}\times\varPhi\times\mathcal{B}$,
where $\mathcal{V}^{s}=\{0,\frac{1}{N_{1}}v_{\max}^{s},\frac{2}{N_{1}}v_{\max}^{s},\ldots,v_{\max}^{s}\}$
is the set of the UAV's speed, $\varPhi=\{0,\frac{1}{N_{2}}2\pi,\frac{2}{N_{2}}2\pi,\ldots,2\pi\}$
is the set of the UAV's direction, and $\mathcal{B}=\{0,1,\ldots,N\}$
is the set of the scheduling actions.}\textcolor{blue}{{} }Then, the
joint action of all UAVs is then given by $\tilde{\bm{a}}(t)=(\bm{a}_{1}(t),\ldots,\bm{a}_{M}(t))$. 

\textcolor{black}{Due to kinematic constraints, not all movements
in $\mathcal{V}^{s}\times\varPhi$ are valid at each state and the
available actions vary by states. To filter out the invalid actions
and prevent unnecessary exploration, we define the set of available
movements for UAV $m$ as $\mathcal{V}'_{m}(t)\triangleq\{v_{m}^{s}(t+1),\varphi_{m}(t)|v_{m}^{s}(t+1)\in\mathcal{V}^{s},\varphi_{m}(t)\in\varPhi,(\varphi_{m}(t-1)-\triangle\varphi_{\max})\textrm{ mod }2\pi\leq\varphi_{m}(t)\leq(\varphi_{m}(t-1)+\triangle\varphi_{\max})\textrm{ mod }2\pi\}$.
 However, due to the long-term constraints in (\ref{eq:energy})-(\ref{eq:start_stop_ps}),
the UAVs may not be able to choose actions from $\mathcal{V}'_{m}(t)$
for the entire period of $T$ time slots. If $\phi_{m}(t)<0$ or $\psi_{m}(t)<0$
, the UAV will not be able to reach the destination within the $T$-th
time slot due to either a lack of time or a lack of energy. In these
cases, the UAV may need to stop exploring and follow predetermined
policies when there is a low time or energy difference. Let $\overline{E}$
denote the maximum energy that can be consumed in one time slot. Since
$\phi_{m}(t+1)\geq\phi_{m}(t)-4$ and $\psi_{m}(t+1)\geq\psi_{m}(t)-4\overline{E}$
(as proven in the appendix), the UAV is free to choose movement actions
from $\mathcal{V}'_{m}(t)$ when $\phi_{m}(t)>4$ and $\psi_{m}(t)>4\overline{E}$.
On the other hand, when $\phi_{m}(t)\leq4$ or $\psi_{m}(t)\leq4\overline{E}$,
there are two predetermined policies: one is that UAV $m$ must adjust
its direction towards the destination and fly at maximum speed if
$\triangle\varphi_{m}^{\textrm{stop}}(t)\leq\triangle\varphi_{\max}\textrm{ or }v_{m}^{s}(t)=0$.
The other is that UAV $m$ must decelerate its speed to zero, accelerate
to the maximum speed, and then fly to the destination with the maximum
speed if $\triangle\varphi_{m}^{\textrm{stop}}(t)>\triangle\varphi_{\max}$
and $v_{m}^{s}(t)\neq0$.}

\textcolor{blue}{}At the same time, UAV $m$ can either schedule
an SN within its coverage area to update or fly without data collection.
However, due to the energy causality, not all of the SNs in the coverage
area of the UAV can be scheduled to transmit. Therefore, the available
scheduling set of UAV $m$ in time slot $t$ is given by 
\begin{equation}
\mathcal{B}'_{m}(t)=\left\{ n\mid n\in\mathcal{B},\overline{d}_{n,m}(t)\leq R_{\textrm{U}},E_{n}(t)\geq E_{n}^{c}\right\} .\label{eq:avail_schedule}
\end{equation}

\subsubsection{State Transition}

\textcolor{black}{We describe the transition of each element of $\bm{s}(t)$
in detail. The battery level and the AoI of each SN are updated as
in (\ref{eq:energySN_update}) and (\ref{eq:update-AOI}), respectively.
The dynamics of the location of UAV $m$ can be expressed as 
\begin{equation}
\bm{u}_{m}(t+1)=\boldsymbol{u}_{m}(t)+(\frac{v_{m}^{s}(t)+v_{m}^{s}(t+1)}{2}\cos\varphi_{m}(t)\tau_{0},\frac{v_{m}^{s}(t)+v_{m}^{s}(t+1)}{2}\sin\varphi_{m}(t)\tau_{0}).\label{eq:location_update}
\end{equation}
 The time difference $\phi_{m}(t)$ is updated based on the location
and velocity of UAV $m$. According to the definition of time difference,
 the update  of the time difference is given by
\begin{equation}
\phi_{m}(t+1)=\phi_{m}(t)-1+T_{m}^{\text{req}}(t)-T_{m}^{\text{req}}(t+1).\label{eq:time_diff_up}
\end{equation}
Similarly, the energy difference is updated as follows:
\begin{equation}
\psi_{m}(t+1)=\psi_{m}(t)-E_{m}(t)+E_{m}^{\textrm{req}}(t)-E_{m}^{\textrm{req}}(t+1).\label{eq:energy_diff_up}
\end{equation}
}

\subsubsection{Cost}

The cost depends on the state and joint action of all UAVs. The overall
goal is to minimize the total AoI. Moreover, if any two UAVs collide,
violating the constraint in (\ref{eq:collision}), a penalty is imposed
on the cost. Then, the cost is defined as follows:
\begin{equation}
c(t)=\begin{cases}
\sum_{n=1}^{N}\delta_{n}(t)+k_{1}, & \textrm{if }\ensuremath{d_{m,m'}(t)<d_{\textrm{safe}},m\neq m',m,m'\in\mathcal{M},}\\
\sum_{n=1}^{N}\delta_{n}(t), & \textrm{otherwise.}
\end{cases}\label{eq:reward_UAVs}
\end{equation}
where $k_{1}$ is a positive constant that is set to be large enough. 

Our goal is to find an optimal policy $\pi^{*}$, which determines
the sequential actions over a finite horizon of length $T$. Since
the system state is partially observed by the UAVs, the policy of
each UAV is a mapping from its observation space to its action space.
In particular, given the initial state $\bm{s}(1)$, the optimal policy
can be obtained by minimizing the time-average expected cost as follows,
\begin{equation}
\pi^{*}=\arg\underset{\pi}{\min}\frac{1}{T}\mathbb{E}_{\pi}\left[\sum_{t=1}^{T}c(t)|s(1)\right].
\end{equation}
Here, the expectation is taken with respect to the distribution over
trajectories induced by $\pi$, along  with the state transitions.\textcolor{black}{{}
However, classic reinforcement learning methods, such as Q-learning,
are not feasible to use when the state and action spaces are large.
This is because one of the conditions for Q-learning to converge to
the optimal solution is to visit each state-action pair an infinite
number of times, which is not possible when the state-action pair
dimension is enormous \cite{Q_learning}.} Therefore, we resort to
multi-agent DRL approach in the following subsection to solve this
problem.

\subsection{Multi-Agent DRL Approach}

If the state of the environment is fully observable by a central controller,
the problem of data collection using multiple UAVs can be solved through
centralized learning. In particular, all the UAVs are controlled by
a single agent, where the input is the state of the environment and
the output is the combination of all UAVs' actions. In centralized
learning, the global action-value function $Q_{\pi}^{G}(\boldsymbol{s}(t),\tilde{\boldsymbol{a}}(t))$,
which is based on the state $\boldsymbol{s}$ and the joint action
$\tilde{\boldsymbol{a}}$, can be expressed as follows:\cite{sutton2018reinforcement}
\begin{equation}
Q_{\pi}^{G}(\boldsymbol{s}(t),\tilde{\boldsymbol{a}}(t))=\mathbb{E}\left[\sum_{i=t}^{T}c(i)|\boldsymbol{s}(t)=\boldsymbol{s},\tilde{\boldsymbol{a}}(t)=\tilde{\boldsymbol{a}}\right],\label{eq:-4}
\end{equation}
This function represents the accumulated expected cost for selecting
the joint action $\tilde{\boldsymbol{a}}$ in state $\boldsymbol{s}$
then following policy $\pi$ since slot $t$. The action-value function
for the optimal policy $Q^{G*}(\boldsymbol{s}(t),\tilde{\boldsymbol{a}}(t))$
can be estimated by \cite{sutton2018reinforcement}
\begin{align}
Q^{G}(\boldsymbol{s}(t),\tilde{\boldsymbol{a}}(t))= & (1-\alpha)Q^{G}(\boldsymbol{s}(t),\tilde{\boldsymbol{a}}(t))+\alpha\left[c(t)+\min_{\tilde{\boldsymbol{a}}'}Q^{G}(\boldsymbol{s}(t+1),\tilde{\boldsymbol{a}}')\right].
\end{align}
The optimal policy is the one that takes the action which minimizes
the action-value function at each step.

However, a UAV cannot obtain the knowledge of other UAVs\textquoteright{}
actions and observations to estimate the global action-value function.
Therefore,  centralized learning cannot be utilized in this distributed
setup. A natural way to realize decentralized learning is to train
each UAV independently by using IDQN \cite{tampuu2017IDQL}, where
each UAV adopts a DQN approach to approximate its action-value function
by using a deep neural network (DNN). Nonetheless, IDQN may not converge
in the non-stationary multi-agent environment due to simultaneous
learning and exploring. 

\begin{figure*}[!t]
\centering\includegraphics[width=0.8\textwidth]{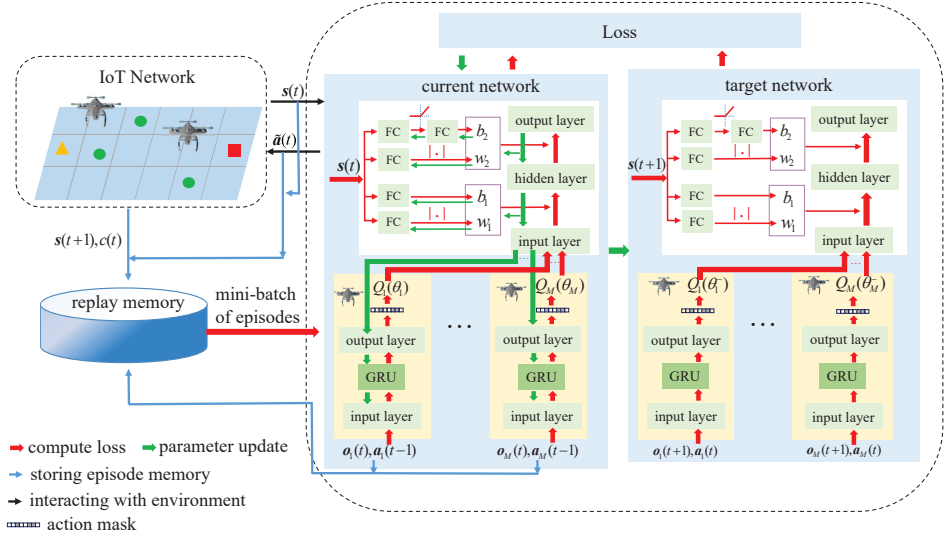}\caption{The framework of the QMIX algorithm in the training phase. The forward
propagation process is represented by a solid red arrow, which is
used to calculate the loss, while the backpropagation process is represented
by a solid green arrow, which is used to update the parameters of
the neural network. The top white box represents the mixing network
and the bottom yellow boxes represent the agent networks. The blue
arrow indicates the collection of the experience, and the black arrow
indicates the interaction with environment. The black grid represents
the action mask. \label{fig:TrainingQMIX}}
\end{figure*}

\begin{figure}
\centering\includegraphics[width=0.6\textwidth]{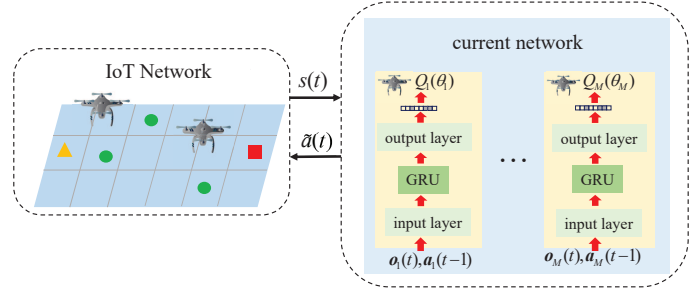}\caption{The framework of the QMIX algorithm in the testing phase.\label{fig:testingQMIX}}
\end{figure}

To address the aforementioned difficulties, we propose a QMIX-based
algorithm, which can be trained in a centralized manner but executed
in a decentralized way \cite{rashid2018qmix}. Similar to IDQN, there
are multiple agent networks in QMIX, each corresponding to a UAV.
In addition, there is a mixing network in QMIX. During the training
process, as shown in Fig. \ref{fig:TrainingQMIX}, each UAV uses an
agent network to estimate the local action-value function based on
its own observations and actions, while the mixing network combines
the system state and all  local action-value functions to produce
a global action-value function. The essence of the QMIX is to guarantee
the monotonicity between the global action-value function and the
local action-value functions. During the testing process, as shown
in Fig. \ref{fig:testingQMIX}, each agent only needs to execute its
action based on its own observations and the previous action, which
will subsequently minimize the global action-value function. In the
following, we explain how QMIX uses the agent networks and the mixing
network to achieve centralized training with decentralized execution.

\subsubsection{Agent Network}

For each UAV, there is an agent network to evaluates its local action-value
function $Q_{m}(\boldsymbol{\tau}_{m}(t),\boldsymbol{a}_{m}(t);\theta_{m})$,
where $\theta_{m}$ is the weights of the agent network. Each agent
network has an input layer, a Gated Recurrent Unit (GRU) layer, and
an output layer. The input layer of the agent network collects the
current observation $\boldsymbol{o}_{m}(t)$ and the previous action
$\boldsymbol{a}_{m}(t-1)$ of UAV $m$ and then feeds the joint observation-action
history $\boldsymbol{\tau}_{m}(t)=(\boldsymbol{o}_{m}(t),\boldsymbol{a}_{m}(t-1))$
to the GRU layer. As an improved version of standard recurrent neural
network, the GRU layer includes update gates and reset gates, which
help capture long-term and short-term dependencies in sequences, respectively
\cite{chung2014GRU}. Hence, the GRU layer is able to overcome the
partial observability of the underlying Dec-POMDP. Then, the GRU layer
outputs its hidden states, which are fed to the output layer. Based
on the hidden states of the GRU layer, the output layer produces the
local action-value function.
\begin{rem}
The computational complexity of training an agent network depends
on the number of neurons in the input layer $h_{\text{I}}^{\text{a}}$,
the GRU layer $h_{\text{G}}^{\text{a}}$, and the output layer $h_{\text{O}}^{\text{a}}$.
Similar to \cite{SakHa_LSTMcomplexAnalyse_2014}, we can obtain the
computational complexity of the GRU layer as $\mathcal{O}(h_{\text{G}}^{\text{a}}(3h_{\text{I}}^{\text{a}}+3h_{\text{G}}^{\text{a}}+h_{\text{O}}^{\text{a}}))$.
Moreover, the computational complexity of a fully-connected layer
with $h_{\text{I}}$ input and $h_{\text{O}}$ output neurons is $\mathcal{O}(h_{\text{I}}h_{\text{O}})$
\cite{Wu_QuantizedConNN_2016_CVPR}. Altogether, the computational
complexity of training an agent network is $\mathcal{O}(h_{\text{I}}^{\text{a}}h_{\text{G}}^{\text{a}}+h_{\text{G}}^{\text{a}}(3h_{\text{I}}^{\text{a}}+3h_{\text{G}}^{\text{a}}+h_{\text{O}}^{\text{a}})+h_{\text{G}}^{\text{a}}h_{\text{O}}^{\text{a}})$,
where $h_{\text{I}}^{\text{a}}$ and $h_{\text{O}}^{\text{a}}$ represent
the dimensions of observation space and action space, respectively. 

\end{rem}

\subsubsection{Mixing Network}

The mixing network consists of a fully-connected neural network (FNN)
with one hidden layer and a set of hypernetworks.  The weights of
the mixing network are denoted by $\theta_{0}$. We use the mixing
network to evaluate the global action-value function $Q^{G}(\tilde{\boldsymbol{\tau}}(t),\tilde{\boldsymbol{a}}(t),$
$\boldsymbol{s}(t);\theta)$, where $\tilde{\bm{\tau}}(t)=(\bm{\tau}_{1}(t),\ldots,\bm{\tau}_{M}(t))$
and $\theta=(\theta_{0},\theta_{1},\ldots,\theta_{M})$. \textcolor{black}{The
input of the mixing network includes the output of each agent network,
i.e., $Q_{m}(\boldsymbol{\tau}_{m},\boldsymbol{a}_{m}),\forall m\in\mathcal{M}$,
and the state of the environment. }In particular, the FNN takes the
output of each agent network as its input and mixes them monotonically
via a hidden layer. The output layer has only one neuron and produces
the global action-value function. Different from a conventional FNN,
the weights and biases of the FNN in the mixing network are produced
by a set of hypernetworks. All the hypernetworks take the state $\boldsymbol{s}(t)$
as input so that $Q^{G}(\tilde{\boldsymbol{\tau}}(t),\tilde{\boldsymbol{a}}(t),\boldsymbol{s}(t);\theta)$
can be adjusted with respect to the state in a non-monotonic way.
There are two hypernetworks that generate the weights between the
input layer and the hidden layer, and those between the hidden layer
and the output layer. Each of these two hypernetworks consists of
one fully-connected (FC) layer. An absolute activation function is
employed at the output of the FC layer to obtain the non-negative
weights of the FNN. Moreover, the biases of the hidden layer and the
output layer of the FNN are produced by another hypernetworks in a
similar manner. In particular, we employ a hypernetwork with one FC
layer to obtain the biases of the hidden layer, and a hypernetwork
containing two FC layers with a rectified linear unit (ReLU) nonlinearity
to obtain the biases of the output layer, respectively.\footnote{Note that the biases are not restricted to being non-negative, as
the monotonicity does not depend on the signs of the biases.} With this structure, the mixing network guarantees that the global
action-value function monotonically increases with a local action-value
function \cite{rashid2018qmix}, i.e., 
\begin{equation}
\frac{\partial Q^{G}(\tilde{\boldsymbol{\tau}}(t),\tilde{\boldsymbol{a}}(t),\boldsymbol{s}(t);\theta)}{\partial Q_{m}(\boldsymbol{\tau}_{m}(t),\boldsymbol{a}_{m}(t);\theta_{m})}\geq0,\forall m\in\mathcal{M}.\label{eq:monotonicity}
\end{equation}
As a result, the joint action that minimizes the global action-value
function is equivalent to the individual actions that minimize each
local action-value function. Specifically, 
\begin{equation}
\arg\min_{\tilde{\boldsymbol{a}}}Q^{G}(\tilde{\boldsymbol{\tau}},\tilde{\boldsymbol{a}},\boldsymbol{s};\theta)=\left(\begin{array}{c}
\arg\underset{\boldsymbol{a}_{1}}{\min}Q_{1}(\boldsymbol{\tau}_{1},\boldsymbol{a}_{1};\theta_{1})\\
\arg\underset{\boldsymbol{a}_{2}}{\min}Q_{2}(\boldsymbol{\tau}_{2},\boldsymbol{a}_{2};\theta_{2})\\
\vdots\\
\arg\underset{\boldsymbol{a}_{M}}{\min}Q_{M}(\boldsymbol{\tau}_{M},\boldsymbol{a}_{M};\theta_{M})
\end{array}\right).\label{eq:IGM-2}
\end{equation}
This allows each agent to greedily choose the action that minimizes
its local action-value function in a decentralized way. 
\begin{rem}
The computational complexity of training the mixing network is dependent
on the number of neurons in the FNN and those in the hypernetworks.
Particularly, the computational complexity of the FNN with one hidden
layer is $\mathcal{O}(h_{\text{I}}^{\text{m}}h_{\text{H}}^{\text{m}}+h_{\text{H}}^{\text{m}}h_{\text{O}}^{\text{m}})$,
where $h_{\text{I}}^{\text{m}}$ is the number of neurons in the input
layer, $h_{\text{H}}^{\text{m}}$ is the number of neurons in the
hidden layer, and $h_{\text{O}}^{\text{m}}$ is the number of neurons
in the output layer. Let $h_{\text{HI}}^{\text{m}}$ denote the number
of neurons of the input layer of the hypernetworks. The computational
complexity for two hypernetworks that generate the weights and biases
between the input layer and the hidden layer are $\mathcal{O}(h_{\text{HI}}^{\text{m}}Mh_{\text{H}}^{\text{m}})$
and $\mathcal{O}(h_{\text{HI}}^{\text{m}}h_{\text{H}}^{\text{m}})$,
respectively. Similarly, the computational complexity for two hypernetworks
that generate the weights and bias between the hidden layer and the
output layer are $\mathcal{O}(h_{\text{HI}}^{\text{m}}h_{\text{H}}^{\text{m}}h_{\text{O}}^{\text{m}})$
and $\mathcal{O}(h_{\text{HI}}^{\text{m}}h_{\text{H}}^{\text{m}}+h_{\text{H}}^{\text{m}}h_{\text{O}}^{\text{m}})$,
respectively. Altogether, the computational complexity of the mixing
network is 
\begin{align}
\mathcal{O}^{\textrm{m}}= & \mathcal{O}(h_{\text{I}}^{\text{m}}h_{\text{H}}^{\text{m}}+h_{\text{H}}^{\text{m}}h_{\text{O}}^{\text{m}}+h_{\text{HI}}^{\text{m}}Mh_{\text{H}}^{\text{m}}+h_{\text{HI}}^{\text{m}}h_{\text{H}}^{\text{m}}+h_{\text{HI}}^{\text{m}}h_{\text{H}}^{\text{m}}h_{\text{O}}^{\text{m}}+h_{\text{HI}}^{\text{m}}h_{\text{H}}^{\text{m}}+h_{\text{H}}^{\text{m}}h_{\text{O}}^{\text{m}})\nonumber \\
= & \mathcal{O}(h_{\text{H}}^{\text{m}}(h_{\text{I}}^{\text{m}}+2h_{\text{O}}^{\text{m}}+(M+2+h_{\text{O}}^{\text{m}})h_{\text{HI}}^{\text{m}})).
\end{align}
Note that $h_{\text{HI}}^{\textrm{m}}$, $h_{\text{I}}^{\text{m}}$,
and $h_{\text{O}}^{\text{m}}$ represent the dimensions of the state
space, the number of agents $M$, and $1$, respectively.

\end{rem}

\subsubsection{Centralized Training}

The centralized training of the agent networks and the mixing network
is conducted offline. Similar to DQN, two sets of neural networks
are used to stabilize the training process, as shown in Fig. \ref{fig:TrainingQMIX}.
One is the current network with weights $\theta$ and the other is
the target network with weights $\theta^{-}$. Here, the weights $\theta=(\theta_{0},\theta_{1},\ldots,\theta_{M})$
include the parameters of the mixing network $\theta_{0}$ and the
parameters of all agent networks $\theta_{m}$ for $m\in\mathcal{M}$.
The current network is used as a function approximator and its weights
are updated at every slot. The target network calculates the target
action-value function, and its weights are fixed for a period of time
before being replaced with the latest weights from the current network
at every O steps.  Moreover, experience replay is also utilized.
In order to efficiently train the GRU layer in each agent network,
we sample the entire experience in an episode, rather than sampling
transition tuples randomly in the reply memory as in DQN. This allows
the hidden states of the GRU layer to learn the temporal correlations
from the experience in the episode \cite{Hausknecht2015DRQN}. 
\begin{algorithm}[tbph]
\caption{Training process of the multi-UAV enabled data collection algorithm.\label{alg:Training-process}}
\begin{algorithmic}[1]

\STATE Initialize the current network $Q^{G}(\tilde{\boldsymbol{\tau}},\tilde{\boldsymbol{a}},\boldsymbol{s};\theta)$
and the target network $Q^{G}(\tilde{\boldsymbol{\tau}},\tilde{\boldsymbol{a}},\boldsymbol{s};\theta^{-})$;

\STATE Initialize the replay memory $D$;

\FOR{ $episode=1:EP$ }

\STATE The environment initializes the initial state $\boldsymbol{s}(1)$
and the agent initializes its observation $\bm{o}_{m}(1)$; 

\WHILE{ $\boldsymbol{s}(t)$ is not a terminal state}

\FOR{ $\textrm{agent}\textrm{ }m=1:M$ }

\STATE \textcolor{black}{Compute available actions $\mathcal{A}_{m}'(t)\triangleq\left(\mathcal{V}_{m}'(t),\mathcal{B}_{m}'(t)\right)$;}

\STATE Select a random action $\boldsymbol{a}_{m}(t)\in\mathcal{A}_{m}'(t)$
with probability $\epsilon$;

\STATE Otherwise select $\boldsymbol{a}_{m}(t)=\arg\min\limits _{\boldsymbol{a}_{m}\in\mathcal{A}_{m}'(t)}Q_{m}(\bm{\tau}_{m}(t),\boldsymbol{a}_{m};\theta_{m}(t))$;

\ENDFOR

\STATE Execute the joint action $\tilde{\boldsymbol{a}}(t)$, and
get cost $c(t)$, next state $\boldsymbol{s}(t+1)$, and next joint
observation $\tilde{\boldsymbol{o}}(t+1)$;

\STATE Store the $(\boldsymbol{s}(t),\tilde{\boldsymbol{\tau}}(t),\tilde{\mathcal{A}}'(t),\tilde{\boldsymbol{a}}(t),c(t),\boldsymbol{s}(t+1),\tilde{\boldsymbol{\tau}}(t+1),\tilde{\mathcal{A}}'(t+1))$
in the replay memory;

\STATE $\boldsymbol{s}(t)=\boldsymbol{s}(t+1)$  ;

\ENDWHILE

\STATE Sample a random mini-batch of episodes from $D$;

\STATE Calculate the target action-value of the episode experiences
$y^{G}$:

\FOR{ $i=1:Z$}

\IF{ $\boldsymbol{s}(i)$ is not the terminal state }

\STATE $y^{G}(i)=c(i)+\underset{\tilde{\boldsymbol{a}}\in\tilde{\mathcal{A}}'(i+1)}{\min}Q^{G}(\tilde{\boldsymbol{\tau}}(i+1),\tilde{\boldsymbol{a}},\boldsymbol{s}(i+1);\theta^{-});$

\ELSE

\STATE $y^{G}(i)=c(i)$;

\ENDIF

\ENDFOR

\STATE $y^{G}=\mathop{\sum_{i=1}^{Z}y^{G}(i)}$;

\STATE Update the current network by performing the gradient descent
in (\ref{eq:Update-QMIX});

\IF{ There are $O$ update-interval steps}

\STATE Synchronize the target network by setting $\theta^{-}=\theta$;

\ENDIF

\ENDFOR

\end{algorithmic}
\end{algorithm}

The current network of QMIX is trained by minimizing the loss function
at each episode. Specifically, the loss function with respect to the
experiences in the episodes sampled from the replay memory is given
by 
\begin{equation}
L(\theta)=\mathop{\frac{1}{Z}\sum_{\textrm{\ensuremath{i=1}}}^{Z}}(y^{G}(i)-Q^{G}(\tilde{\boldsymbol{\tau}}(i),\tilde{\boldsymbol{a}}(i),\boldsymbol{s}(i);\theta))^{2},\label{eq:Loss-qmix-1}
\end{equation}
where $Z$ is the number of transitions in the mini-batch and $y^{G}(i)=c(i)+\underset{\tilde{\boldsymbol{a}}\in\tilde{\mathcal{A}}'(i+1)}{\min}Q^{G}(\tilde{\boldsymbol{\tau}}\left(i+1\right),$
$\tilde{\boldsymbol{a}},\boldsymbol{s}(i+1);\theta^{-})$. Here, $\tilde{\mathcal{A}}'(i+1)=(\mathcal{A}'_{1}(i+1),\ldots,\mathcal{A}'_{M}(i+1))$
is the joint available actions of all UAVs in time slot $i+1$, and
$\mathcal{A}'_{m}(i+1)$ is the available actions of UAV $m$ in time
slot $i+1$ obtained by the action mask method based on the observation
$\boldsymbol{o}_{m}(i+1)$. Note that $Q^{G}(\tilde{\boldsymbol{\tau}}(i),\tilde{\boldsymbol{a}}(i),\boldsymbol{s}(i);\theta)$
and $Q^{G}(\tilde{\boldsymbol{\tau}}(i+1),\tilde{\boldsymbol{a}},\boldsymbol{s}(i+1);\theta^{-})$
are obtained by the current network and the target network, respectively.
Afterward, the weights of the current network are updated by the semi-gradient
algorithm as follows:
\begin{align}
 & \theta=\theta+\frac{\alpha}{Z}\sum_{\textrm{\ensuremath{i=1}}}^{Z}(y^{G}(i)-Q^{G}(\tilde{\boldsymbol{\tau}}(i),\tilde{\boldsymbol{a}}(i),\boldsymbol{s}(i);\theta))\nabla_{\theta}Q^{G}(\tilde{\boldsymbol{\tau}}(i),\tilde{\boldsymbol{a}}(i),\boldsymbol{s}(i);\theta).\label{eq:Update-QMIX}
\end{align}

 The details of the centralized training are summarized in Algorithm
\ref{alg:Training-process}. First, the current network and target
network are initialized and the replay memory is cleared out. Since
the current network is updated based on the episodes of experiences,
we gather the experience from an entire episode (Lines 5\textasciitilde 14)
before executing a gradient descent step (Line 25). Since the action
mask filters out invalid actions which violate constraints (\ref{eq:speed_constra})-(\ref{eq:start_stop_ps}),
there are two cases making an episode terminate: 1) when two UAVs
collide, $d_{m,m'}(t)<d_{\textrm{safe}},m\neq m',m,m'\in\mathcal{M}$;
2) all the UAVs reach their final positions at time slot $T$. Before
the terminal state, each agent chooses an action by utilizing the
$\varepsilon$-greedy policy simultaneously, where $\varepsilon$
is decreasing with the increasing number of time slots $t$ (Lines
6\textasciitilde 10). After all the agents conduct their actions,
a cost $c(t)$ and the next observation $\tilde{\boldsymbol{o}}(t+1)$
can be obtained. The environment also transits to the next state $\boldsymbol{s}(t+1)$
(Line 11). \textcolor{blue}{{} }Then, we randomly sample a mini-batch
of episodes from the replay memory $D$ to update the current network
(Lines 15\textasciitilde 25). In every fixed $O$ episodes, the target
network is updated by copying the current network parameters (Lines
26\textasciitilde 28). \textcolor{black}{Note that the number and
position of SNs remain constant throughout all episodes.}

\subsubsection{Distributed Execution}

When the offline centralized training is finished, the agent networks
can be executed online in a decentralized manner. Particularly, each
UAV make decisions in a distributed way by leveraging the pre-trained
agent network based on its own observations. The testing framework
of QMIX is shown in Fig. \ref{fig:testingQMIX}. Specifically, in
every slot, each UAV chooses the action that minimizes its local action-value
function, i.e., $\bm{a}_{m}(t)=\arg\min\limits _{\boldsymbol{a}_{m}\in\mathcal{A}_{m}'(t)}Q_{m}(\boldsymbol{\tau}_{m},\boldsymbol{a}_{m};\theta_{m})$,
and updates its observations accordingly. Each UAV continues its flight
until it reaches its final position at the $T$-th slot. The details
of the distributed execution are summarized in Algorithm \ref{alg:Testing-process}.
\begin{algorithm}[!tbph]
\caption{Testing process of the multi-UAV enabled data collection algorithm.\label{alg:Testing-process}}

\begin{algorithmic}[1]

\STATE Initialize the environment;

\FOR{ $episode=1:EPT$ }

\STATE The environment initializes the initial state $\boldsymbol{s}(1)$
and the agent initializes its observation $\bm{o}_{m}(1)$;

\WHILE{ $\boldsymbol{s}(t)$ is not the terminal state}

\FOR{ $\textrm{agent}\textrm{ }m=1:M$ }

\STATE\textcolor{blue}{{} }\textcolor{black}{Compute available actions
$\mathcal{A}_{m}'(t)$;}

\STATE Select $\boldsymbol{a}_{m}(t)=\arg\min\limits _{\boldsymbol{a}_{m}\in\mathcal{A}_{m}'(t)}Q_{m}(\bm{\tau}_{m}(t),\boldsymbol{a}_{m};\theta_{m}(t))$;

\ENDFOR

\STATE Execute the joint action $\tilde{\boldsymbol{a}}(t)$, and
get cost $c(t)$, next state $\boldsymbol{s}(t+1)$, and next joint
observation $\tilde{\boldsymbol{o}}(t+1)$;

\STATE $\boldsymbol{s}(t)=\boldsymbol{s}(t+1)$;

\ENDWHILE

\ENDFOR

\end{algorithmic}
\end{algorithm}

\section{Simulation Results \label{sec:Simulation-Results}}

In this section, we conduct extensive simulations to evaluate the
performance of our proposed algorithm. First, we describe the simulation
setup and three baseline algorithms. Then, we evaluate the convergence
and effectiveness of the proposed algorithm with respect to different
system parameters.

\subsection{Simulation Setup}

In the simulation, we deploy the SNs uniformly at random in a square
area of $800\,\textrm{m}\times800\,\textrm{m}$.  Without loss of
generality, we  set the initial and final positions of the UAVs evenly
at the bottom and top of the area, respectively. For example, if $M=3$,
the UAVs\textquoteright{} initial locations will be  $(0\,\textrm{m},0\,\textrm{m},100\,\textrm{m})$,
$(360\,\textrm{m},0\,\textrm{m},100\,\textrm{m})$, and $(760\,\textrm{m},0\,\textrm{m},100\,\textrm{m})$,
and the final locations will be  $(0\,\textrm{m},760\,\textrm{m},100\,\textrm{m})$,
$(360\,\textrm{m},760\,\textrm{m},100\,\textrm{m})$, and $(760\,\textrm{m},760\,\textrm{m},100\,\textrm{m})$.
It is worth noting that the initial and final positions of each UAV
can be the same, and the initial/final positions of different UAVs
can also be the same. All the UAVs and the SNs are fully charged at
the beginning of each episode. Unless otherwise stated, the simulation
parameters of the IoT network can be found in Table \ref{tab:system-parameters}.

\begin{table}
\caption{System parameters\label{tab:system-parameters}}

\centering

\begin{tabular}{c|c|c|c}
\hline 
\textbf{Parameter} & \textbf{Value} & \textbf{Parameter} & \textbf{Value}\tabularnewline
\hline 
Flight altitude $z$ & $100\,$m \cite{WFY_UAV-device-multiDRL_TCOM_2021} & Air density $\rho$ & $1.225\,$kg/$\textrm{\ensuremath{\textrm{m}^{3}}}$\tabularnewline
\hline 
Time duration $T$ & 100 slots & Disc area for each rotor $A$ & $0.0314\,\textrm{m}^{2}$\tabularnewline
\hline 
Slot length $\tau_{0}$ & $0.5\,$s & Transmit power of each SN $P_{c}$ & $5\,$mW \cite{leng2019age}\tabularnewline
\hline 
Safe distance $d_{\textrm{safe}}$ & $10\,$m & Noise power $\sigma^{2}$ & $-110\,$dBm \cite{HHM_AoI-traj-plan-data-collec-wireless-power-IoT2021}\tabularnewline
\hline 
UAV maximum speed $v_{\max}^{s}$ & $20\,$m/s \cite{RDing_3DTra-Freq-EE-FairC-DRL_2020TWC} & $\beta_{0}$, $\beta_{1}$ & 11.95, 0.14 \cite{mozaffari2017mobile}\tabularnewline
\hline 
UAV maximum turning angle $\triangle\varphi_{\max}$ & $\pi/3$ & $\eta_{LoS}$, $\eta_{NLoS}$ & $1.6\,$dB, $23\,$dB \tabularnewline
\hline 
Battery capacity  $E_{\text{max}}$,$E_{\textrm{max}}^{\textrm{sn}}$ & $2.4\times10^{4}\,$J, $5\,$mJ & Carrier frequency$f_{c}$ & $2\,$GHz \cite{mozaffari2017mobile}\tabularnewline
\hline 
$E_{\textrm{n}}^{\textrm{har}}$, $\lambda_{\textrm{n}}$ & $0.42\,$mJ, $0.9$ & Light speed in a vacuum $c$ & $3\times10^{8}\,$m/s\tabularnewline
\hline 
UAV mass $W$ & $2\,$kg \cite{RDing_3DTra-Freq-EE-FairC-DRL_2020TWC} & Number of rotors $n_{r}$ & 4\tabularnewline
\hline 
Gravity acceleration $g$ & $9.8\,$m/$\textrm{s}^{2}$ & $c_{f},c_{T}$ & 0.131, 0.302\tabularnewline
\hline 
Local blade section drag coefficient $\sigma$ & 0.012 & $c_{s},d_{0}$ & 0.0955, 0.834\tabularnewline
\hline 
\end{tabular}
\end{table}

The QMIX-based algorithm is implemented by the PyTorch Library. Specifically,
the agent network includes one GRU layer with $256$ hidden neurons.
Since each UAV has the same state space and action space, the agent
network of each UAV is identical \cite{foerster2016parameter_sharing}.
  The hidden layer in the mixing network contains $256$ neurons.
The hyperparameters of the QMIX-based algorithm are listed in Table
\ref{tab:Hyperparameter-of-QMIX-based}.

\begin{table}[tp]
\centering\caption{Hyperparameters of the QMIX-based algorithm\label{tab:Hyperparameter-of-QMIX-based}}

\begin{tabular}{c|c|c|c}
\hline 
\textbf{Parameter} & \textbf{Value} & \textbf{Parameter} & \textbf{Value}\tabularnewline
\hline 
The number of training episodes $EP$ & 50000 & Minimum $\varepsilon$ & 0.01\tabularnewline
\hline 
 $D,O$ & 1000, 200 & Learning rate $\alpha$ & 0.0005\tabularnewline
\hline 
Mini-batch size & 32 & Optimizer & Adam\tabularnewline
\hline 
Initial $\varepsilon$ & 0.99 & Activation function & ReLU\tabularnewline
\hline 
$\varepsilon$-greedy decrement & 9.9e-6 & $N_{1},N_{2}$ & 1, 6\tabularnewline
\hline 
\end{tabular}
\end{table}

In the following, we compare the proposed QMIX-based algorithm with
three baseline algorithms:
\begin{itemize}
\item Nearest scheduling algorithm: This is a simplified version of the
QMIX-based algorithm, where the trajectory is designed with the same
method as in the multi-agent DRL approach, but each UAV schedules
the SN with the nearest distance to itself.
\item Cluster-based algorithm: In this algorithm, all the SNs are first
clustered into $M$ clusters by K-means, with the initial positions
of the cluster center being the start positions of the UAVs. Then,
each UAV collects data from the SNs in its corresponding cluster.
In each time slot, the UAV flies towards the SN with the largest AoI
in its cluster and schedules the SN with the largest AoI within its
coverage.
\item IDQN-based algorithm: In this algorithm, each UAV works as an independent
agent and employs DQN to address the trajectory design and the SN
scheduling. 
\end{itemize}

\subsection{Performance Evaluation}

\begin{figure}
\subfloat[]{\includegraphics[width=0.5\textwidth]{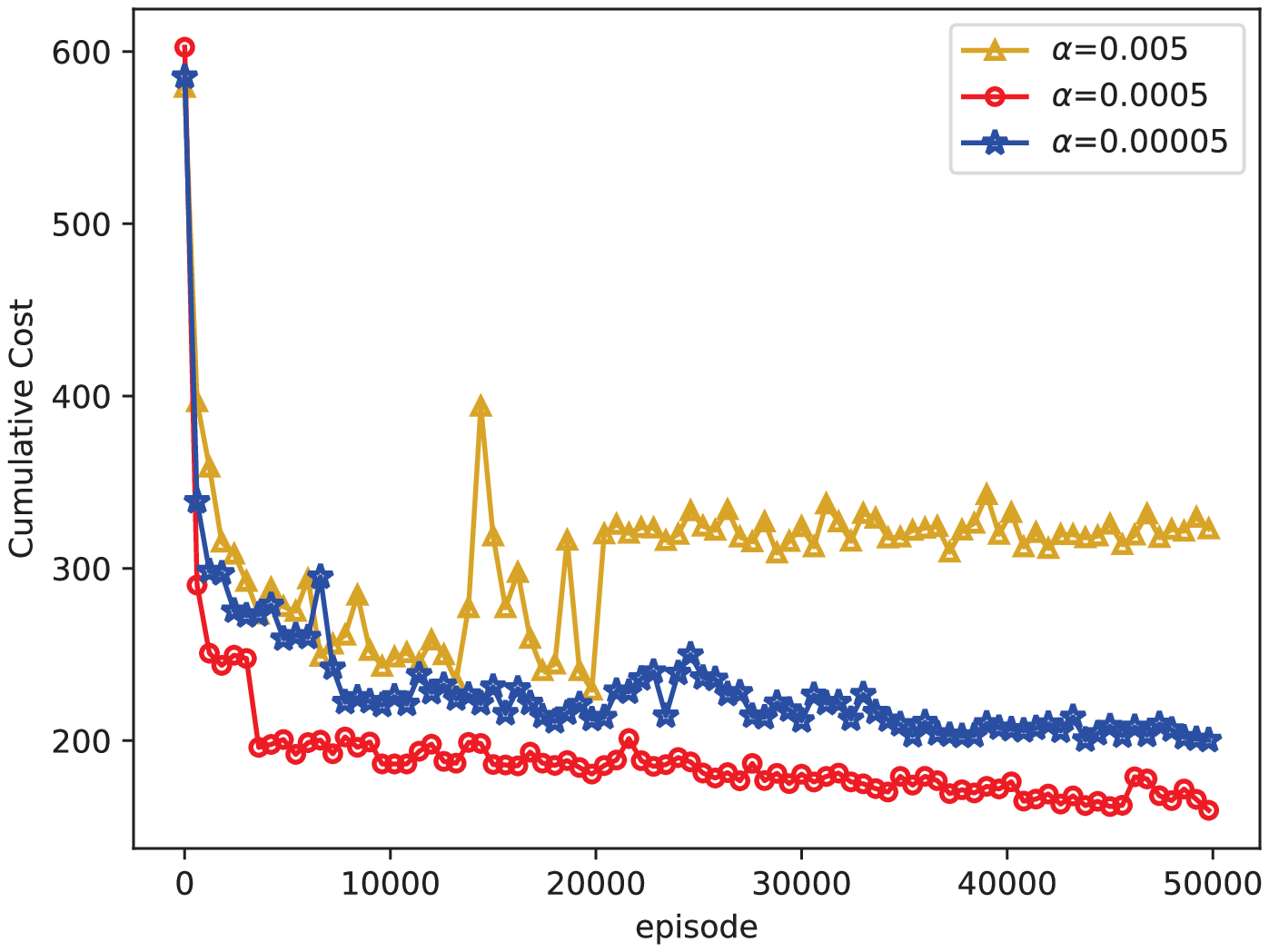}\centering

}\subfloat[]{\includegraphics[width=0.5\textwidth]{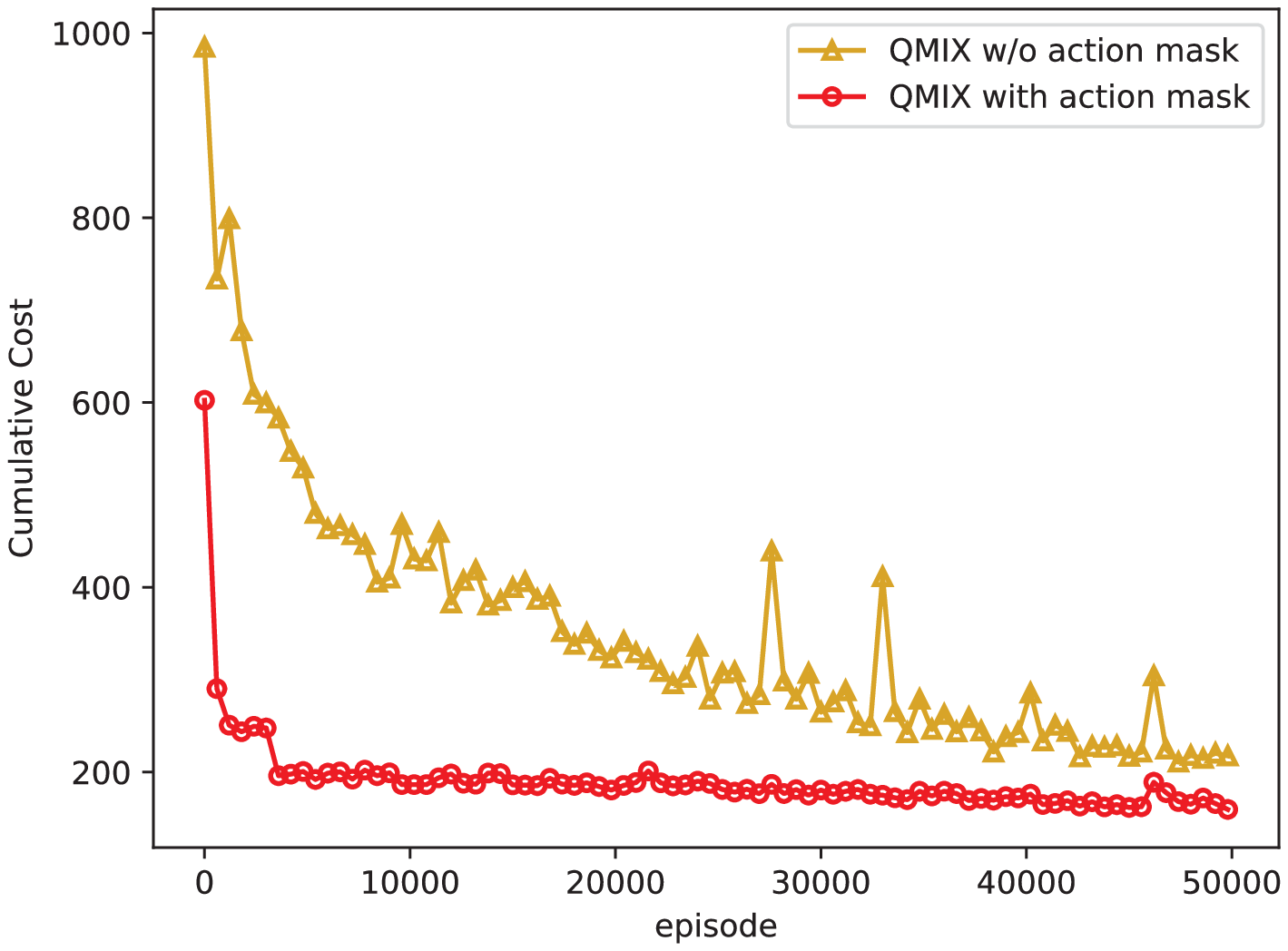}\centering

}

\caption{The convergence evaluation of our proposed QMIX-based algorithm  ($\xi_{\textrm{th}}=5\textrm{ dB}$,
$M=2$, and $N=15$). (a) The influence of different learning rates
$\alpha$.\label{fig:convergence} (b) The influence of the action
mask method.}
\end{figure}

We illustrate the convergence of the proposed QMIX-based algorithm
in Fig. \ref{fig:convergence}(a). In particular, the cumulative cost
is plotted against the number of training episodes for different learning
rates $\alpha$. As seen in the figure, a learning rate that is too
high causes instability, while a rate that is too low leads to convergence
at a local optimum. In this case, we find that the learning rate of
0.0005 performs best, as it produces the lowest cumulative cost compared
to rates of 0.005 and 0.00005. This rate is therefore chosen for use
with the QMIX algorithm in the following analysis. \textcolor{black}{In
Fig. \ref{fig:convergence}(b), we examine the effect of the action
mask method on QMIX's convergence. By comparing QMIX with and without
this method, we can see that the inclusion of the action mask leads
to faster convergence and a lower cumulative cost. This is due to
the mask's ability to prevent invalid explorations by hiding actions
that violate the constraints, thereby improving the performance of
the QMIX algorithm.}

\begin{figure}
\subfloat[]{\centering\includegraphics[width=0.5\textwidth]{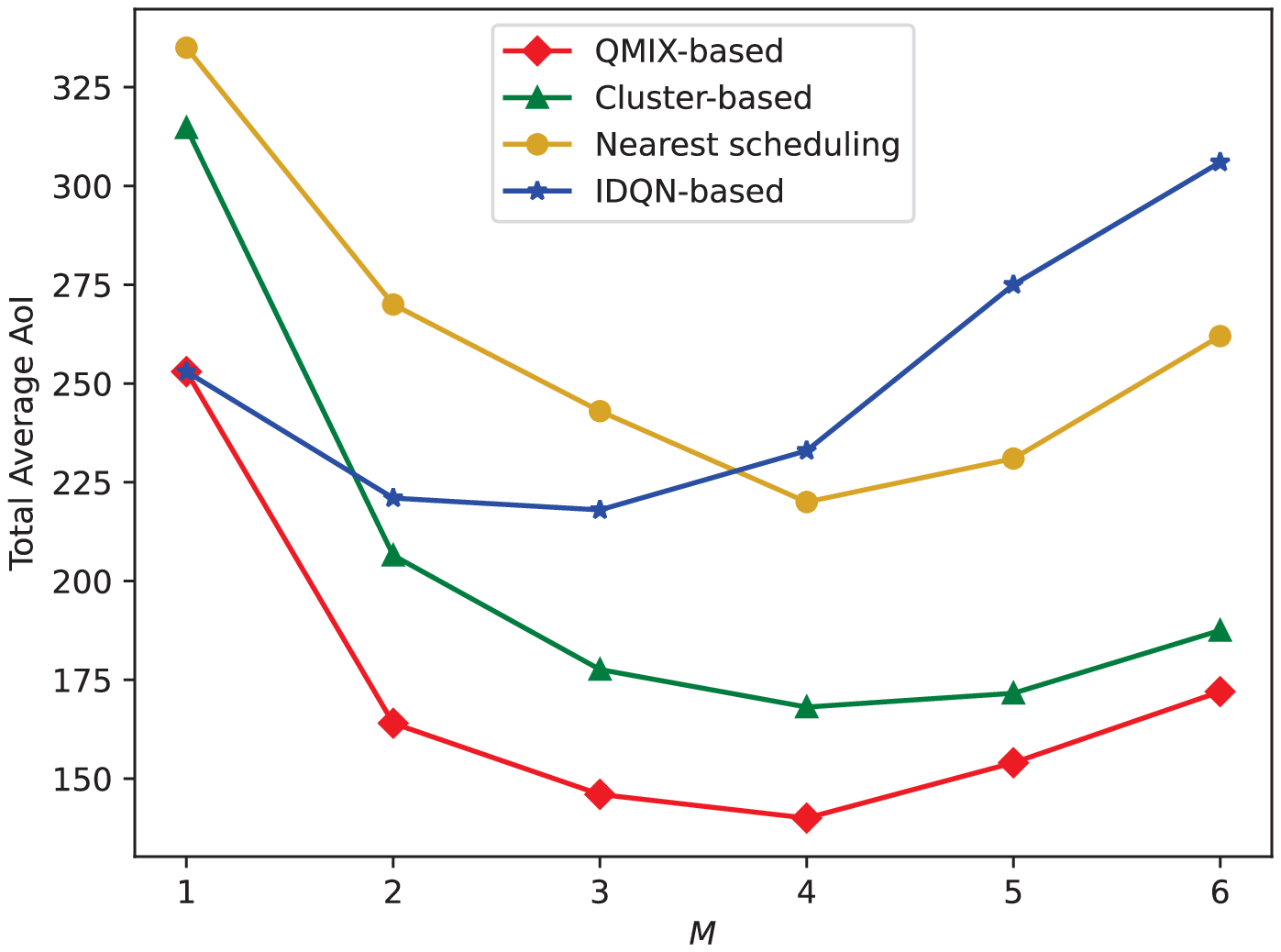}

}\subfloat[]{\includegraphics[width=0.5\textwidth]{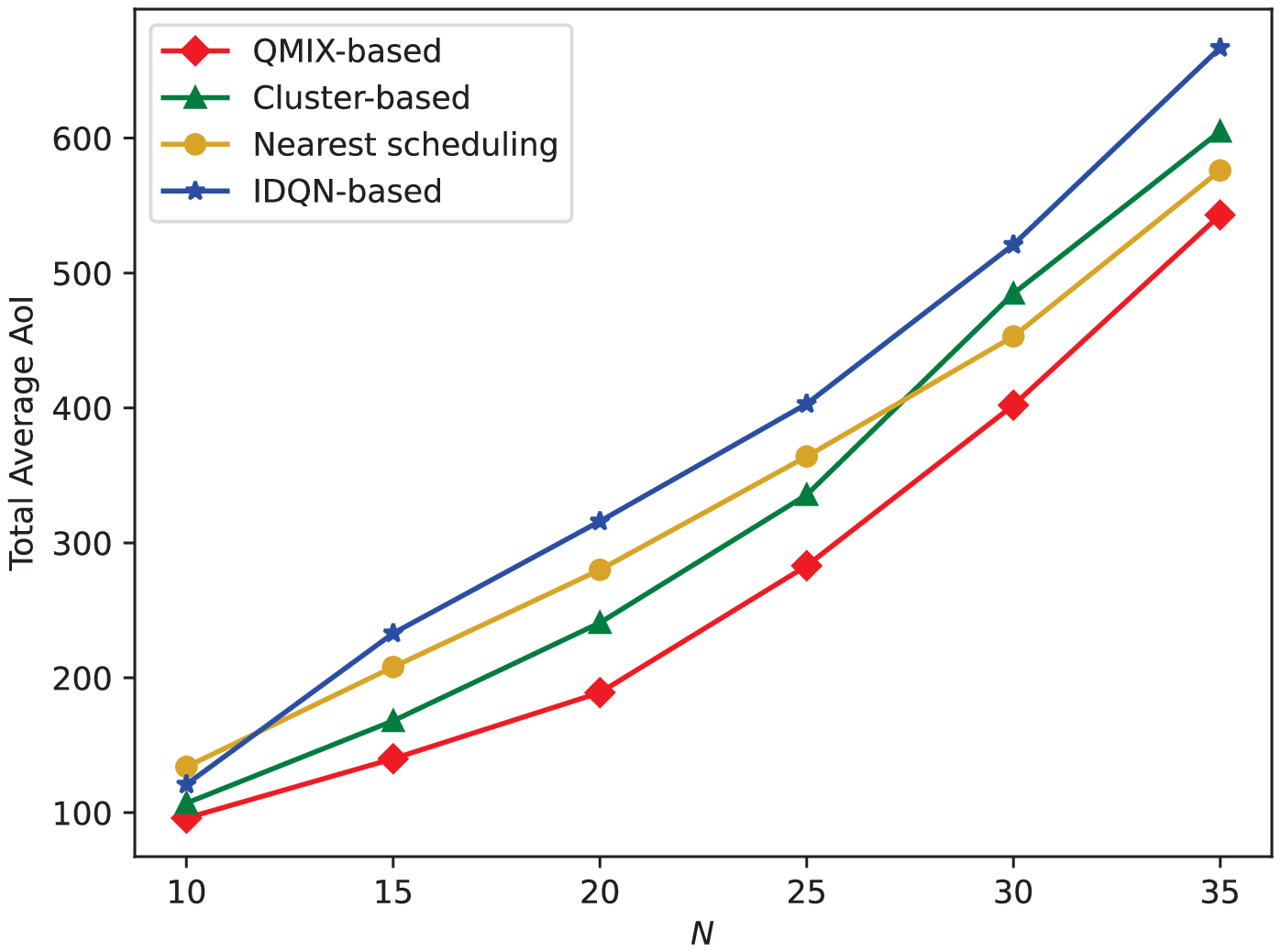}\centering

}

\caption{Comparison between QMIX-based algorithm and other three baseline algorithms
($\xi_{\textrm{th}}=5\textrm{ dB}$). (a) The total average AoI versus
the number of UAVs $M$ with $N=15$.\label{fig:Effect-M} (b) The
total average AoI versus the number of SNs $N$ with $M=4$.}
\end{figure}

Fig. \ref{fig:Effect-M}(a) shows the total average AoI with respect
to the number of UAVs $M$ with $N=15$ and $\xi_{\textrm{th}}=5\textrm{ dB}$.
As $M$ increases, the total average AoI of the QMIX-based algorithm
first decreases and then increases.This trend can be explained from
two aspects. On the one hand, the average number of SNs that each
UAV needs to schedule decreases as $M$ grows. Since each UAV can
schedule no more than one SN at a time, each SN can be scheduled more
frequently with more UAVs, resulting in a decrease in the total average
AoI. On the other hand, the interference level also grows as $M$
increases. Hence, the probability of the status update being successfully
received by the UAV is reduced, thereby raising the total average
AoI. Clearly, the performance of the QMIX-based algorithm coincides
with that of the IDQN-based algorithm when $M=1$, since they both
degrade to single-agent DRL. Fig. \ref{fig:Effect-M}(b) illustrates
the total average AoI versus the number of SNs $N$ with $\xi_{\textrm{th}}=5\textrm{ dB}$
and $M=4$. We can see that the total average AoI increases as $N$
increases. This is because that each UAV can only schedule at most
one SN to update its status at a time. Hence, the larger the number
of SNs, the longer each SN must wait. Moreover, the total average
AoI increases quickly as the number of SNs continues to increase.
This is because, for a large $N$, more SNs have to  wait to be scheduled
by a limited number of UAVs,  resulting in a rapid increase  in the
total average AoI. In both subfigures, the total average AoI of the
proposed QMIX-based algorithm is lower than that of the other three
baseline algorithms. This advantage is achieved by jointly optimizing
the trajectories of UAVs and the scheduling of SNs through cooperative
training of the UAVs with the global value function.

\begin{figure}
\subfloat[]{\includegraphics[width=0.5\textwidth]{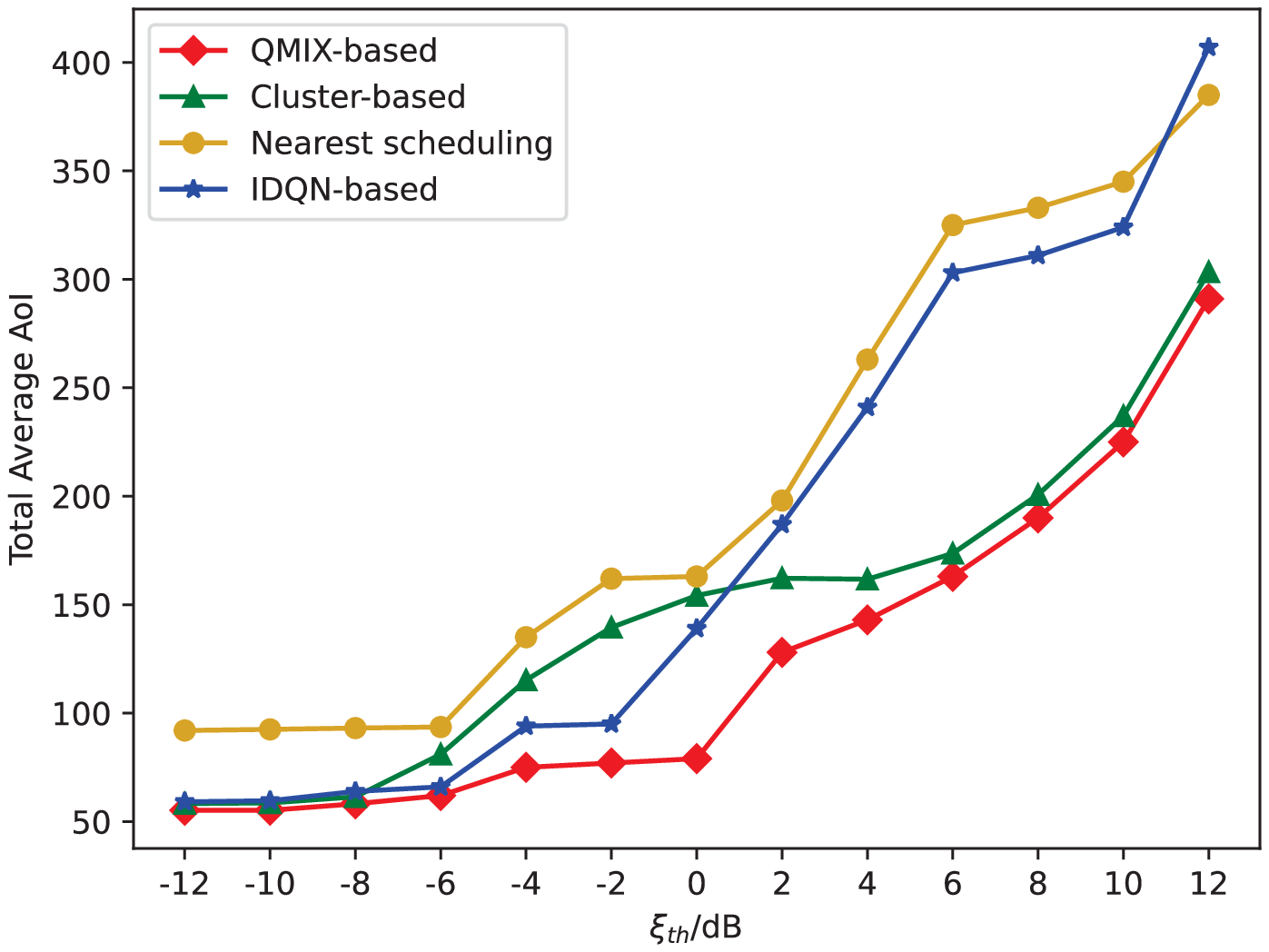}\centering

}\subfloat[]{\includegraphics[width=0.5\textwidth]{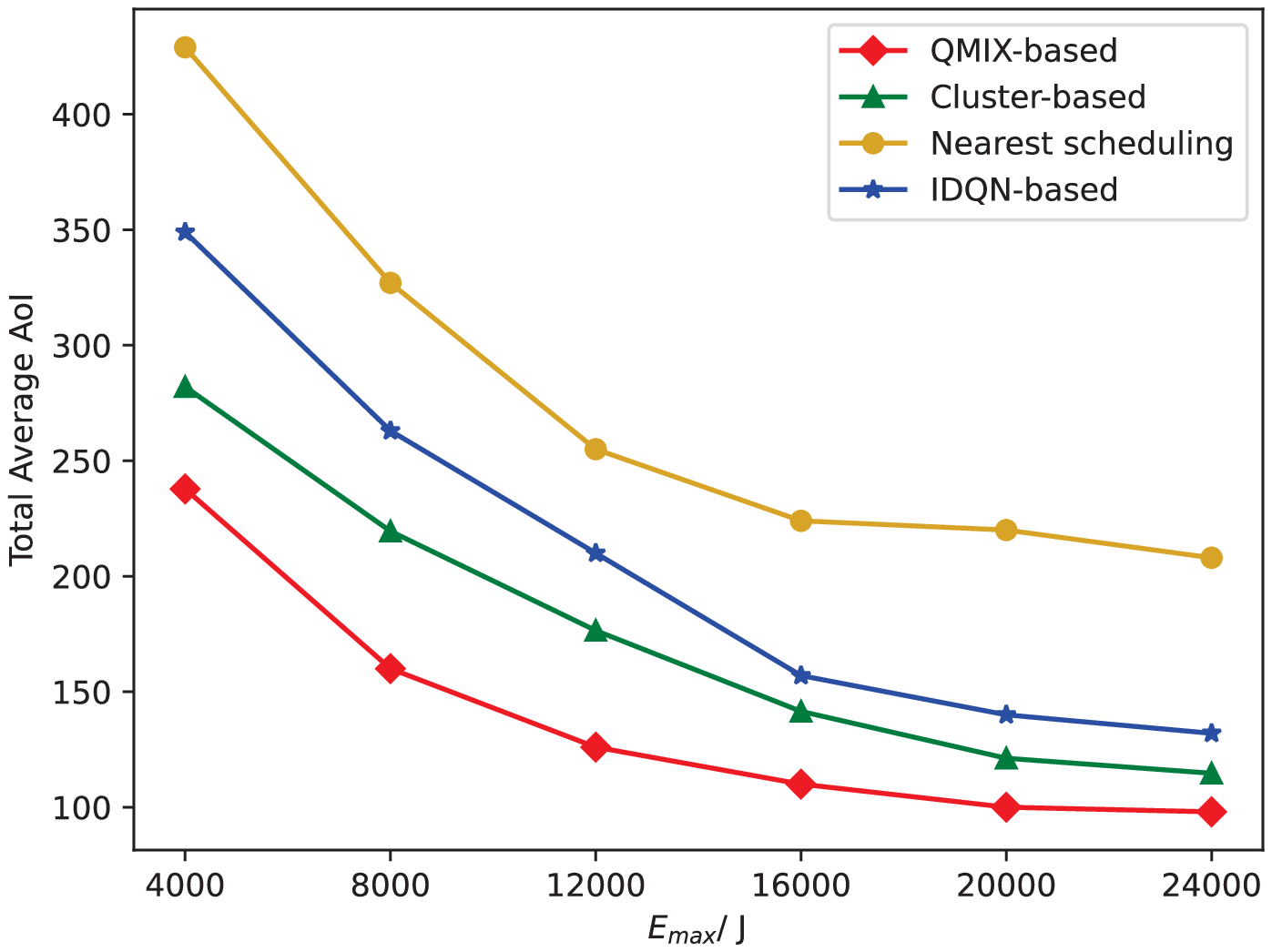}\centering

}\caption{Comparison between QMIX-based algorithm and other three baseline algorithms
($N=15$). (a) The total average AoI versus the SINR threshold $\xi_{\textrm{th}}$
with $M=4$.\label{fig:SINR} (b) The total average AoI versus the
total energy $E_{\textrm{max}}$ with $\xi_{\textrm{th}}=5\textrm{ dB and }M=3$.}
\end{figure}

 Fig. \ref{fig:SINR}(a) shows the total average AoI versus the threshold
$\xi_{\textrm{th}}$ with $N=15$ and $M=4$. From Fig. \ref{fig:SINR}(a),
we can see that the total average AoI increases as $\xi_{\textrm{th}}$
increases. This is due to the fact that the probability of successful
transmission decreases with a larger $\xi_{\textrm{th}}$ , resulting
in a higher total average AoI. It can also be seen that, while the
cluster-based algorithm performs similarly to the QMIX-based algorithm
in both the low and high regimes of $\xi_{\textrm{th}}$, the QMIX-based
algorithm consistently has the lowest total average AoI across all
regimes of $\xi_{\textrm{th}}$.  Fig. \ref{fig:SINR}(b) evaluates
the total average AoI versus the UAV's battery capacity $E_{\textrm{max}}$
with $N=15$, $\xi_{\textrm{th}}=5\textrm{dB}$, and $M=3$. The initial
and final locations are set to the same, with the UAVs starting at
 $(0\,\textrm{m},360\,\textrm{m},100\,\textrm{m})$, $(360\,\textrm{m},360\,\textrm{m},100\,\textrm{m})$,
and $(760\,\textrm{m},360\,\textrm{m},100\,\textrm{m})$, respectively.
The energy consumed by a UAV in one time slot is mainly determined
by the UAV's speed and acceleration, as shown in Eq. (\ref{eq:Energy_Consumption}).
 From Fig. \ref{fig:SINR}(b), we can see that the total average
AoI decreases sharply with $E_{\textrm{max}}$ at first and then decreases
more slowly. The reason is that the energy constraint is tight when
$E_{\textrm{max}}$ is small, requiring the UAV to hover in some slots
to conserve energy. When $E_{\textrm{max}}$ is larger, the energy
constraint is less restrictive, allowing the UAV to move more freely
to collect packets.

\begin{figure}
\includegraphics[width=0.5\textwidth]{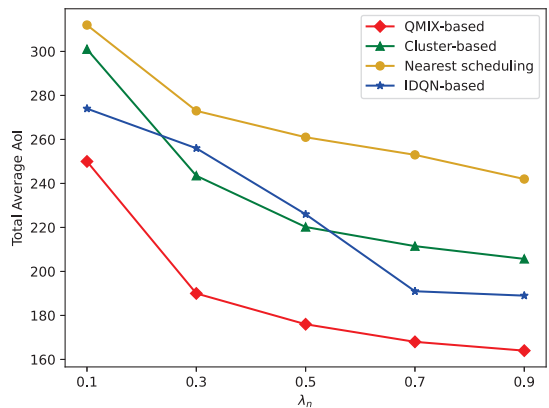}\centering\caption{The total average AoI versus the SN's energy arrival probability $\lambda_{n}$
with $N=15$, $\xi_{\textrm{th}}=5\textrm{dB}$, and $M=2$.\label{fig:lamda_n}}
\end{figure}

 Fig. \ref{fig:lamda_n} shows the effect of SN's energy arrival probability
$\lambda_{n}$ on the total average AoI with $N=15$, $\xi_{\textrm{th}}=5\textrm{dB}$,
and $M=2$. From Fig. \ref{fig:lamda_n}, we can see that as $\lambda_{n}$
increases, the total average AoI decreases. This is because when $\lambda_{n}$
is small, the energy of the SN is insufficient, resulting in untimely
status updates due to the lack of energy. However, as $\lambda_{n}$
increases, the SN is more likely to have enough energy to update when
the UAV flies to its vicinity. When $\lambda_{n}\geq0.7$, the decrease
in total average AoI becomes marginal because the SN can be charged
in time. We can also see  that the QMIX-based algorithm achieves the
lowest total average AoI compared to other baseline algorithms. This
is because the QMIX-based algorithm can jointly optimize UAVs' trajectories
and scheduling strategies, even when $\lambda_{n}$ is small, and
effectively adjust them to reduce the total average AoI.

\begin{figure*}[t]
\subfloat[\label{fig:multi_a}]{\includegraphics[width=0.25\textwidth]{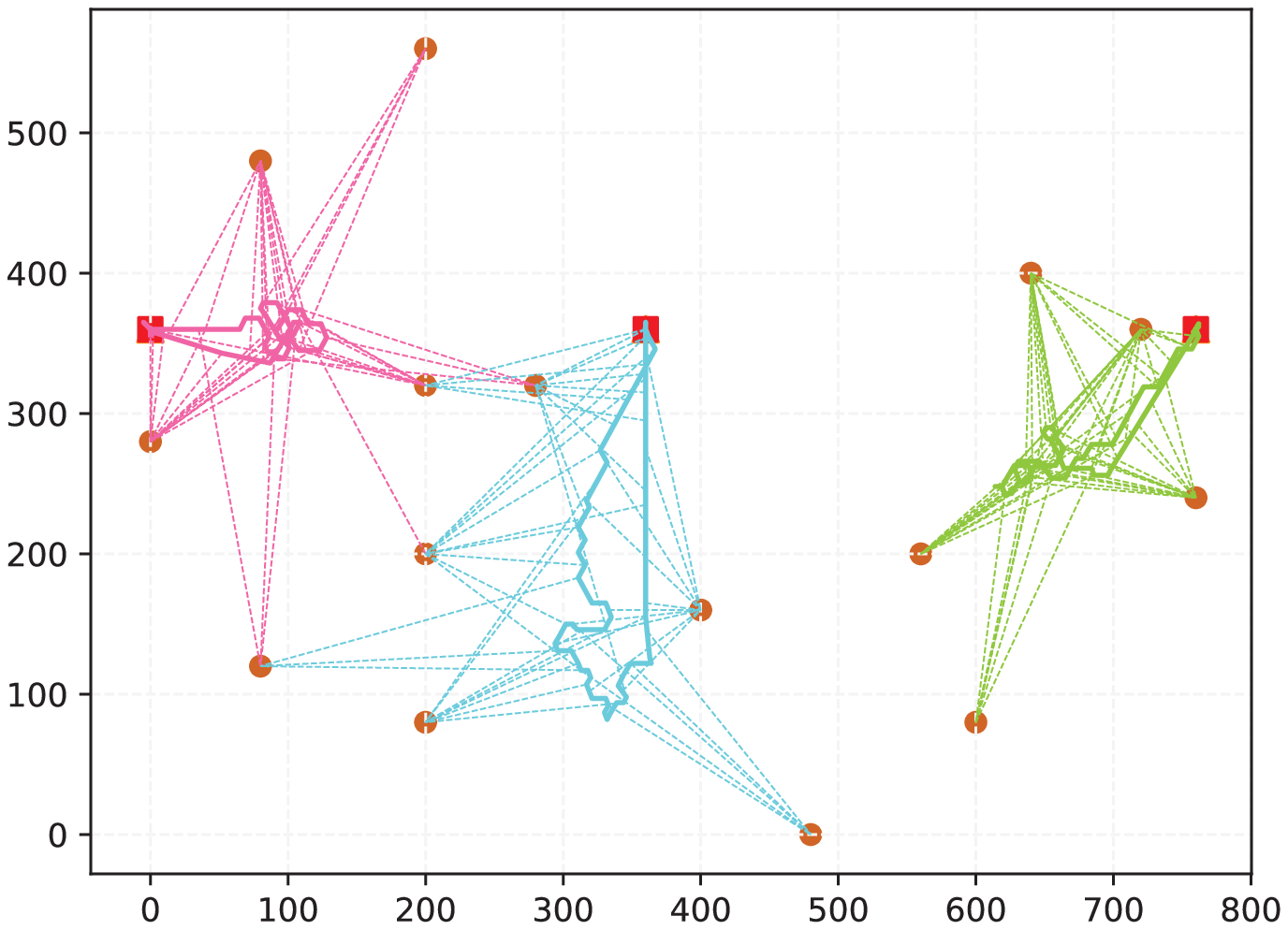}

}\subfloat[\label{fig:multi_b}]{\includegraphics[width=0.25\textwidth]{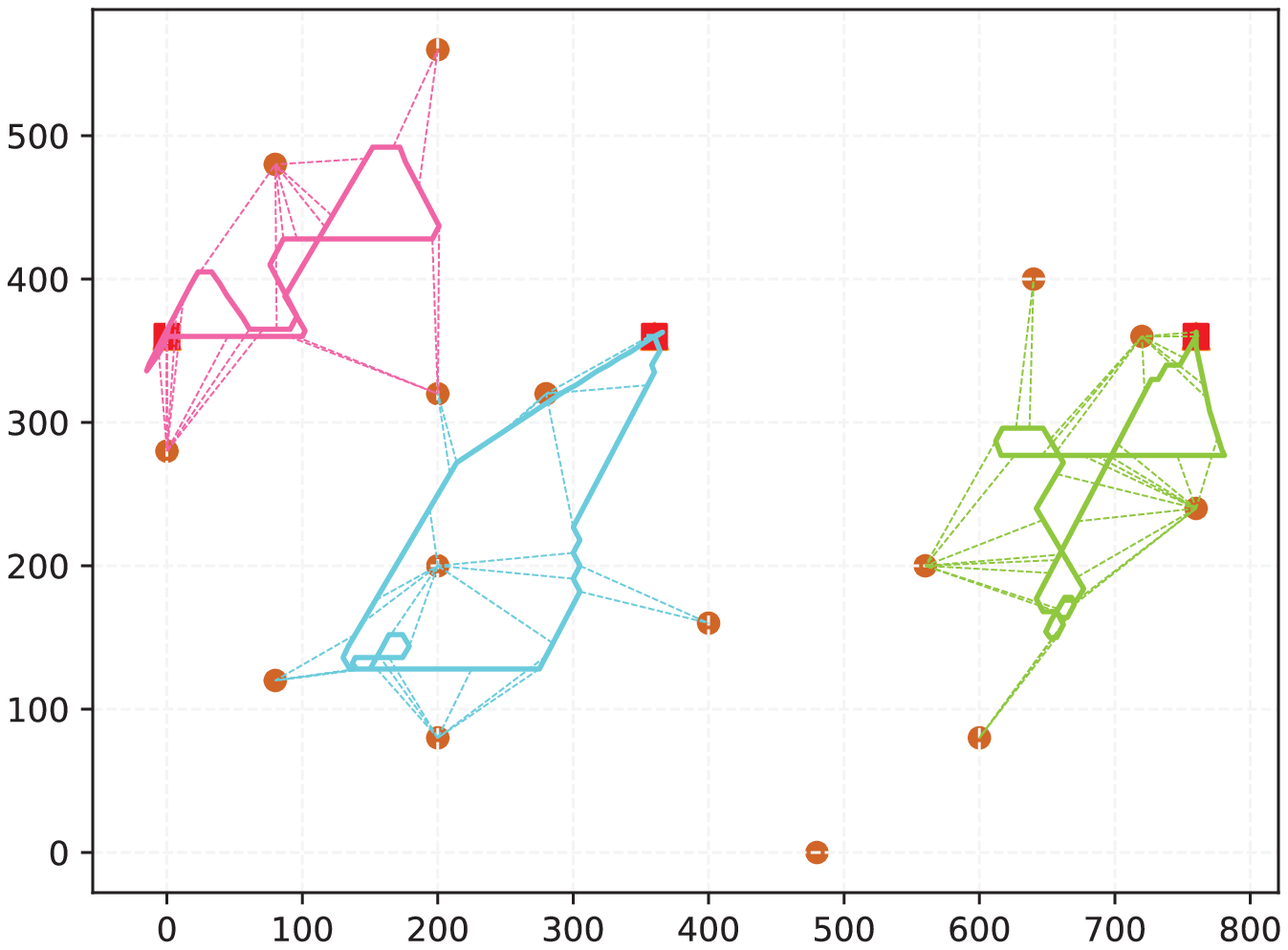}

}\subfloat[\label{fig:multi_c}]{\includegraphics[width=0.25\textwidth]{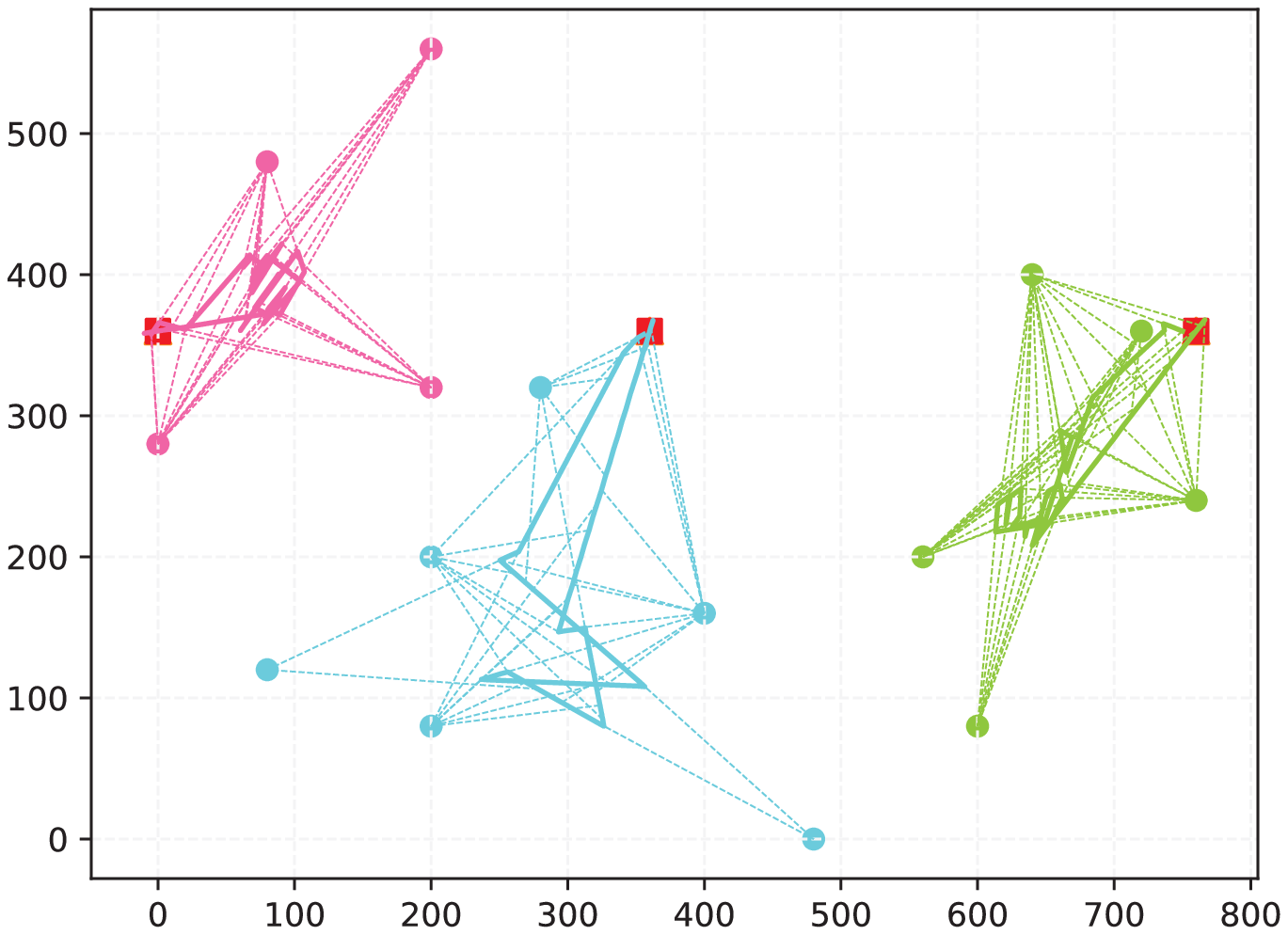}}\subfloat[]{\includegraphics[width=0.25\textwidth]{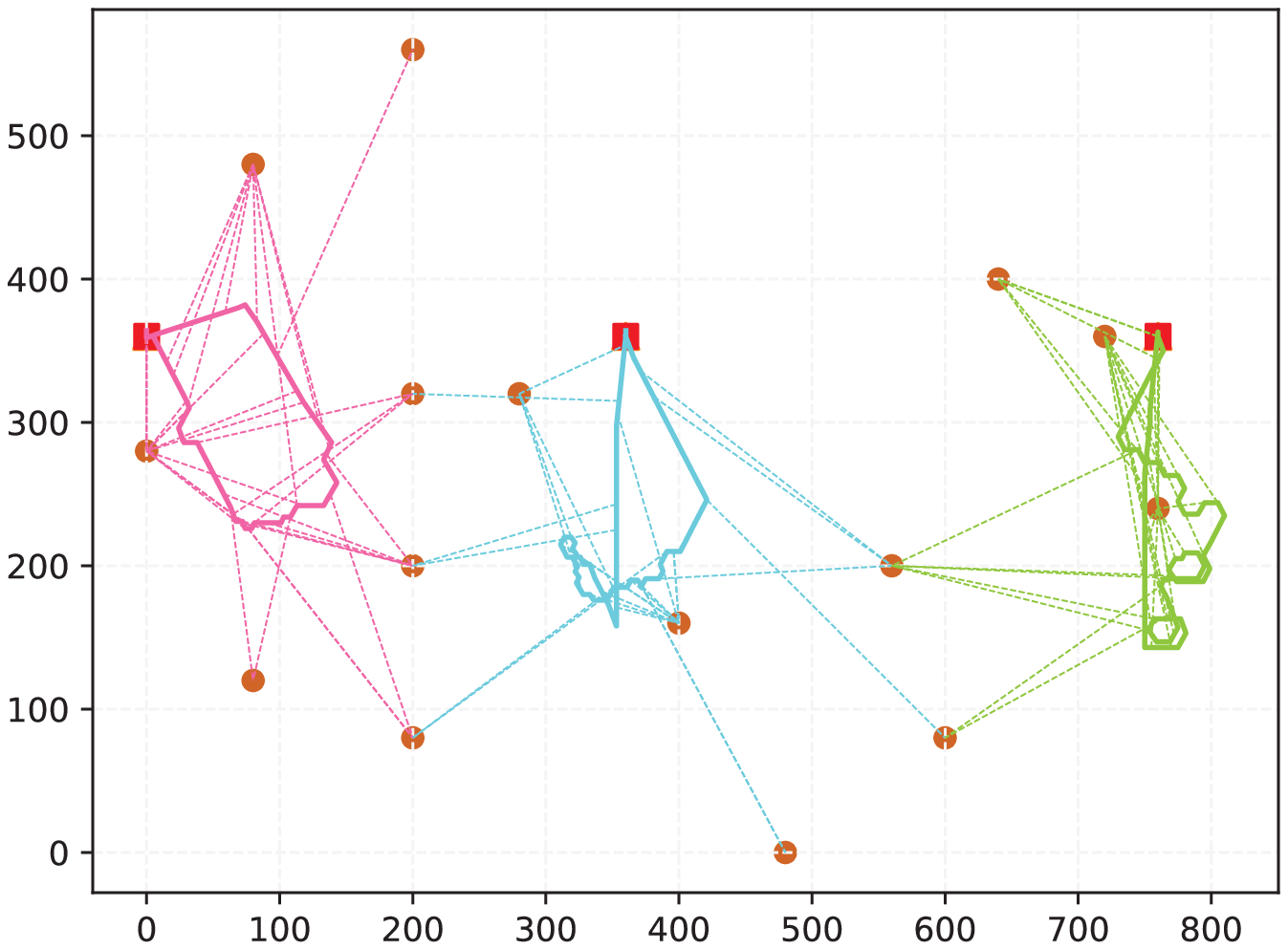}

}\centering

\subfloat[]{\includegraphics[width=0.25\textwidth]{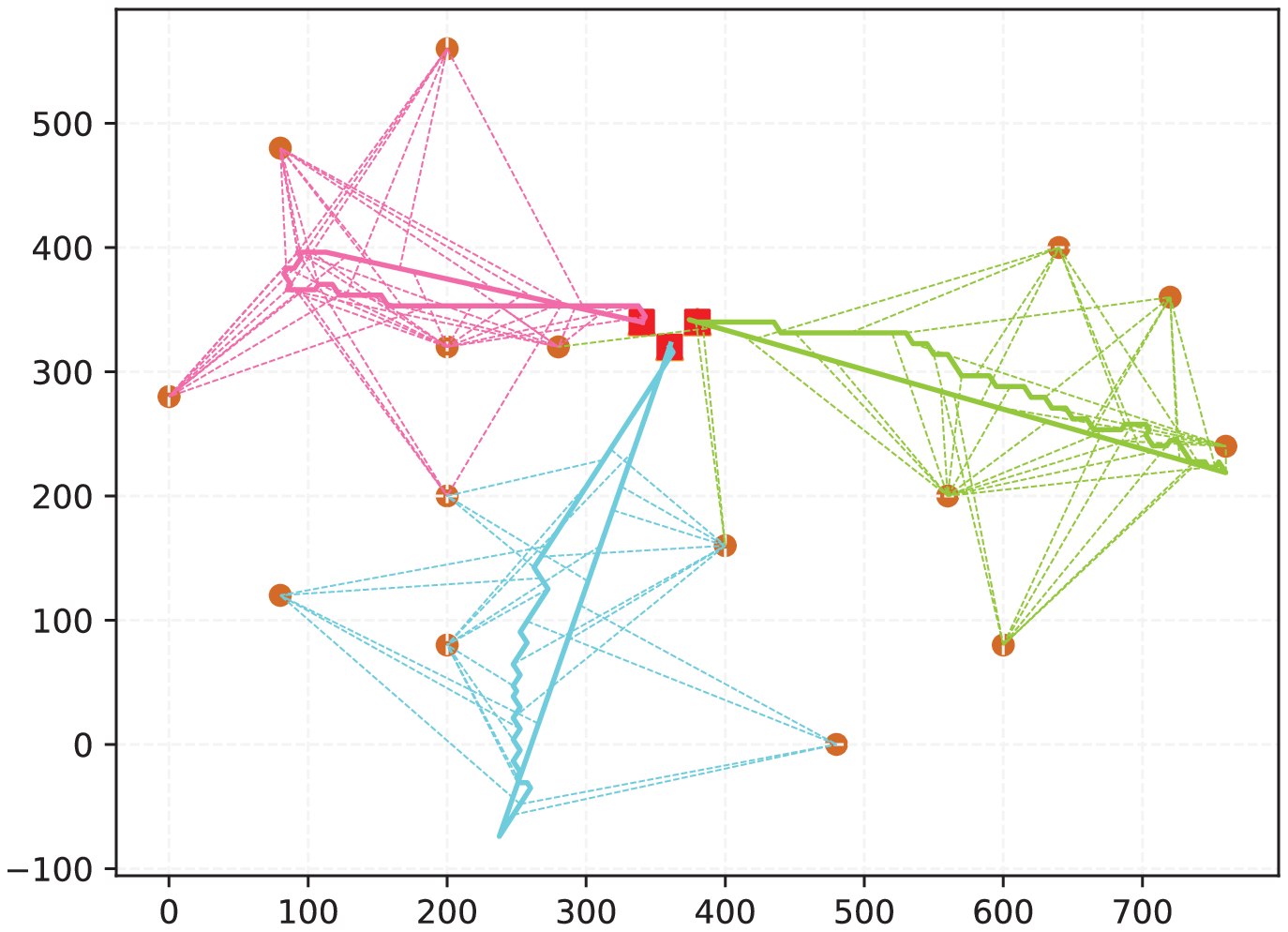}

}\subfloat[]{\includegraphics[width=0.25\textwidth]{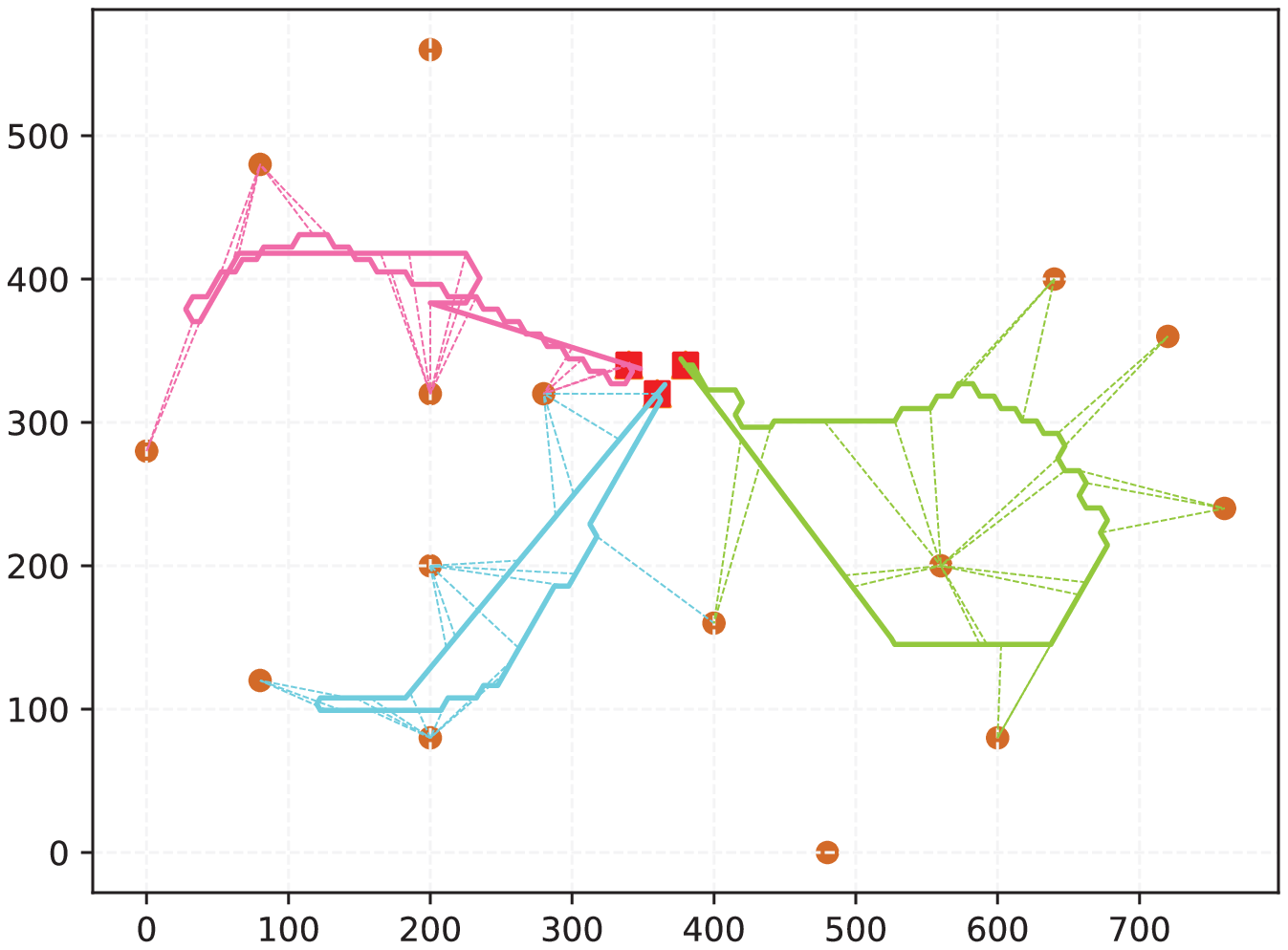}

}\subfloat[]{\includegraphics[width=0.25\textwidth]{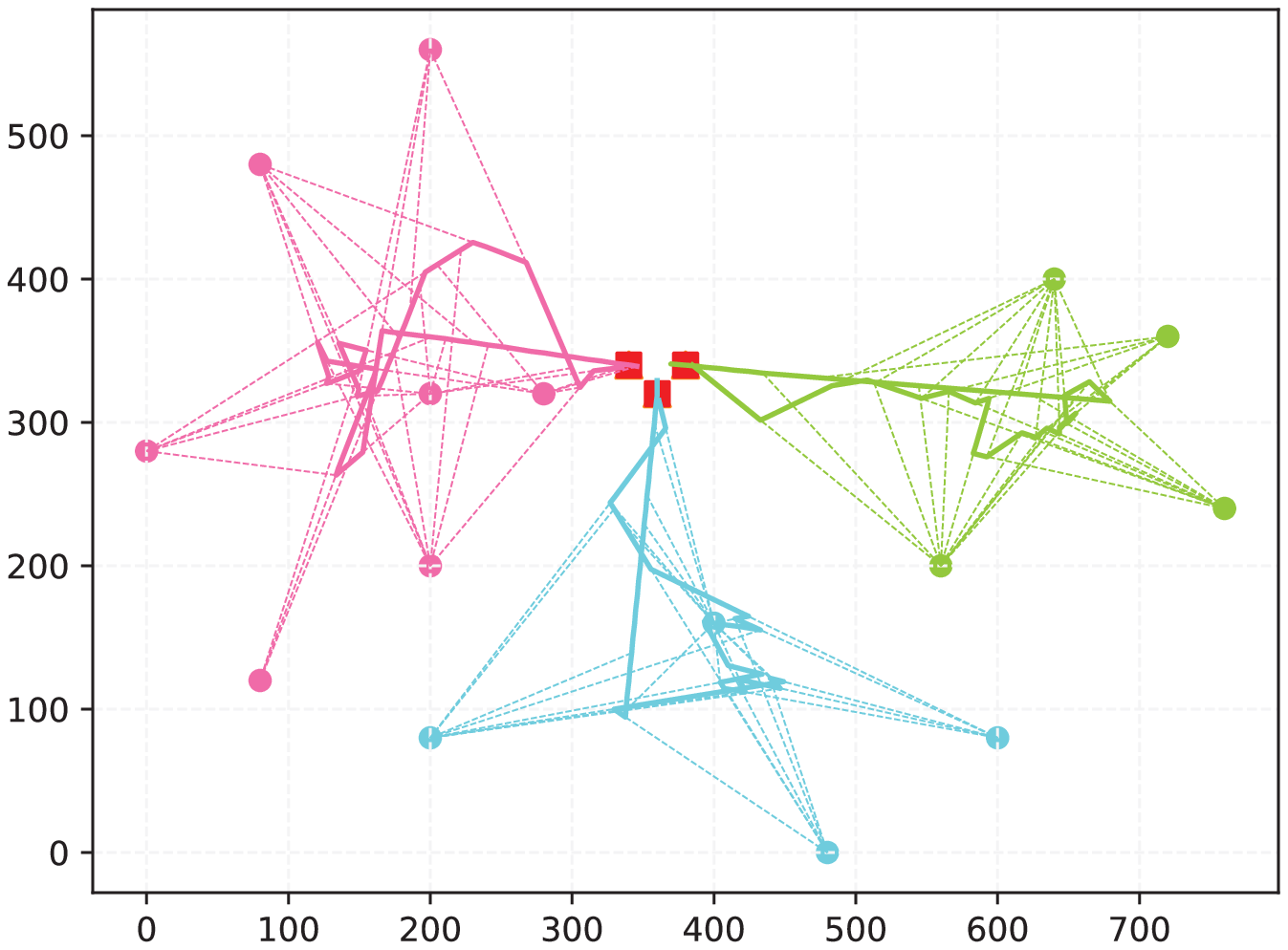}

}\subfloat[]{\includegraphics[width=0.25\textwidth]{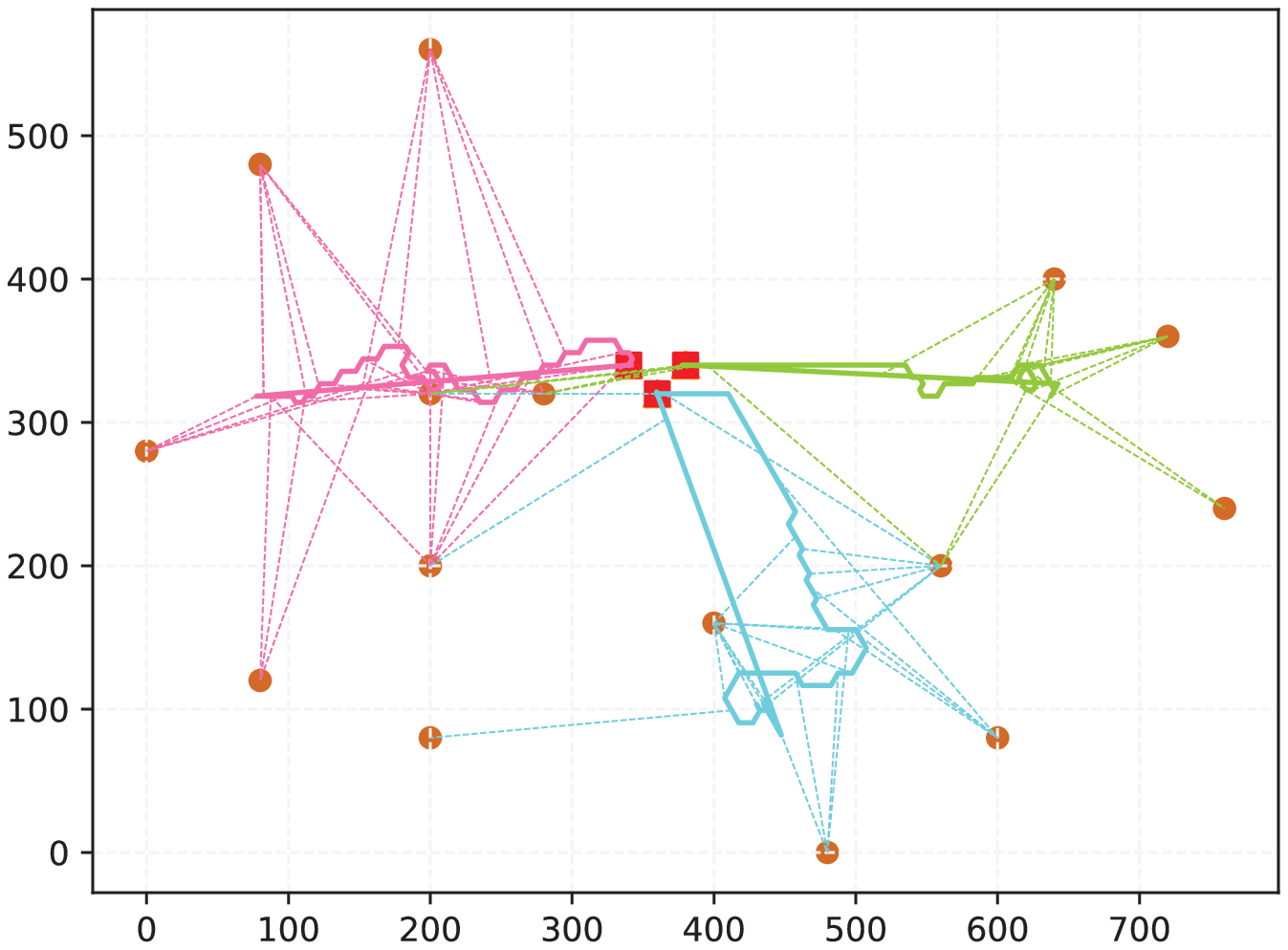}

}

\caption{An illustration of the trajectories of UAVs and scheduling of SNs
in one episode ($\xi_{\textrm{th}}=7\textrm{ dB}$, $N=15$, $p_{h}=0.9$,
and $M=3$). (a)  QMIX-based algorithm with initial and final positions
far apart. (b)  Nearest scheduling algorithm with initial and final
positions far apart. (c)  Cluster-based algorithm with initial and
final positions far apart. (d)  IDQN-based algorithm with initial
and final positions far apart. (e)  QMIX-based algorithm with close
initial and final positions. (f)  Nearest scheduling algorithm with
close initial and final positions. (g)  Cluster-based algorithm with
close initial and final positions. (h)  IDQN-based algorithm with
close initial and final positions.\label{fig:Tra_schedule}}
\end{figure*}

Fig. \ref{fig:Tra_schedule} displays the trajectories of UAVs and
the scheduling of SNs in different algorithms. The solid line indicates
the trajectory of the UAV, while the dashed line represents the scheduling
of SNs. Moreover, we use different colors to denote different UAVs
and their scheduled SNs. For each UAV, its initial and final locations
are set to the same and represented by  squares in the figures. In
Figs. \ref{fig:Tra_schedule}(a)-(d), the initial and final positions
of the UAVs are far apart, while in Figs. \ref{fig:Tra_schedule}(e)-(h),
the initial and final positions are close to each other. In Fig.
\ref{fig:Tra_schedule}(a), the multiple UAVs can fly to and schedule
the SNs cooperatively so as to reduce the interference and the total
average AoI. However, in Fig. \ref{fig:Tra_schedule}(b), the scheduling
of the SNs are not coordinated in the nearest scheduling algorithm,
resulting in higher interference.  In Fig. \ref{fig:Tra_schedule}(c),
the trajectories of the UAVs are suboptimal in the cluster-based algorithm
because the status updates of SNs can only be collected by a specific
UAV once they are clustered. In Fig. \ref{fig:Tra_schedule}(d), the
IDQN-based algorithm can only optimize the trajectories of individual
UAVs and the scheduling of SNs separately, resulting in a local optimal
solution. As shown in Figs. \ref{fig:Tra_schedule}(e)-(h), our approach
can also work well in scenarios where the initial and final locations
of the UAVs are close to each other. Through the comparison, we can
see that the UAVs in the QMIX-based algorithms are able to divide
the workload and collect data from different SNs. Additionally, they
can cooperate to collect data for some SNs, resulting in a reduction
of the total average AoI.

\textcolor{black}{The comparisons of complexity and performance between
our proposed algorithm and other baseline algorithms are summarized
in Table \ref{tab:Complexity}. The  complexity of the Cluster-based
algorithm is primarily determined by the K-means algorithm, which
has a complexity of $\mathcal{O}(MN)$. On the other hand, the  complexity
of the QMIX-based, Nearest Scheduling, and IDQN-based algorithms is
mainly determined by the computation of their neural networks. The
QMIX-based algorithm and the Nearest Scheduling algorithm have the
same agent and mixing networks, but the output layer of the agent
network in the Nearest Scheduling algorithm is smaller. To clearly
illustrate the difference, the computational complexity of the agent
network $\mathcal{O}^{\textrm{a}}$ is denoted as a function of the
output dimension of the agent network $h_{\text{O}}$. The IDQN-based
algorithm has the same agent networks as the QMIX-based algorithm,
but lacks the mixing network. Therefore, the complexity of the QMIX-based
algorithm is higher than that of other baseline algorithms. The table
also shows the total average AoI for all baseline algorithms and our
proposed algorithm, which was obtained with $N=15,M=4,E_{\max}=2.4\times10^{4}\,\textrm{J},\textrm{ and }\xi_{\textrm{th}}=5\,\textrm{dB}$.
 We can see that the QMIX-based algorithm achieves the best performance.}

\begin{table}
\centering

\caption{Computational complexity and performance comparisons\label{tab:Complexity}}

\begin{tabular}{c|c|c}
\hline 
Method & Computational Complexity & Total average AoI\tabularnewline
\hline 
QMIX-based & $M\mathcal{O}^{\textrm{a}}((N+1)(N_{1}+1)(N_{2}+1))+\mathcal{O}^{\textrm{m}}$ & 140\tabularnewline
\hline 
Cluster-based & $\mathcal{O}(MN)$ & 168.1\tabularnewline
\hline 
Nearest scheduling & $M\mathcal{O}^{\textrm{a}}((N_{1}+1)(N_{2}+1))+\mathcal{O}^{\textrm{m}}$ & 208\tabularnewline
\hline 
IDQN-based & $M\mathcal{O}^{\textrm{a}}((N+1)(N_{1}+1)(N_{2}+1))$ & 233\tabularnewline
\hline 
\end{tabular}
\end{table}

\section{Conclusions\label{sec:Conclusions}}

In this study, we examined the problem of optimally collecting data
in multi-UAV enabled IoT networks using multiple cooperative UAVs.
We took into account kinematic, energy, trajectory, and collision
avoidance constraints and aimed to minimize the total average AoI.
To accomplish this goal, we proposed the QMIX-based algorithm to jointly
optimize the trajectories of the UAVs and the scheduling of the SNs.
Particularly, the UAVs were centrally trained using a global value
function, and then each UAV carried out data collection in a distributed
manner based on its local observations.\textcolor{black}{{} Our simulation
results showed that the proposed QMIX-based algorithm was superior
to other baseline approaches and the action mask method helped accelerate
the convergence of the QMIX-based algorithm. We observed that the
UAVs in the QMIX-based algorithm were able to divide their workload
and collect data from different SNs, and in some cases even cooperated
in collecting data from the same SN. We also found that there is an
optimal number of UAVs that minimizes the total average AoI due to
the tradeoff between increased transmission opportunities and higher
mutual interference. In the future work, we will consider optimizing
of trajectory planning transmission scheduling based on the NR frame
structure, where the UAV trajectory can be optimized on a large timescale,
and the transmission scheduling is optimized on a small time scale.}\textcolor{blue}{{}
}

\appendices{}

\section{\textcolor{black}{Proof for The Property of $\phi_{m}(t+1)\protect\geq\phi_{m}(t)-4$
and $\psi_{m}(t+1)\protect\geq\psi_{m}(t)-4\overline{E}$}}

\textcolor{black}{ When the UAV is able to immediately turn its
velocity direction towards the destination at the beginning of slot
$t$, i.e., $\triangle\varphi_{m}^{\textrm{stop}}(t)\leq\triangle\varphi_{\max}\textrm{ or }v_{m}^{s}(t)=0$,
and the direction of velocity at the beginning of slot $t+1$ is opposite
to the direction of the current position pointing towards the destination,
i.e., $\varphi_{m}(t+1)=\varphi_{m}^{\textrm{stop}}(t)+\pi$, the
minimum values of $(\phi_{m}(t+1)-\phi_{m}(t))$ and $(\psi_{m}(t+1)-\psi_{m}(t))$
are obtained. }

\textcolor{black}{In this case, $\bm{u}_{m}(t+1)$, $\bm{u}_{m}(t)$,
and $\bm{u}_{m}^{\text{stop}}$ are on a straight line. Hence, we
have $\left\Vert \bm{u}_{m}(t)-\bm{u}_{m}^{\text{stop}}\right\Vert -\left\Vert \bm{u}_{m}(t+1)-\bm{u}_{m}^{\text{stop}}\right\Vert =-\frac{v_{m}^{s}(t)+v_{m}^{s}(t+1)}{2}\tau_{0}$.
Then, we can obtain 
\begin{align}
 & T_{m}^{\textrm{req}}(t)-T_{m}^{\textrm{req}}(t+1)\nonumber \\
= & \left\lceil \frac{\left\Vert \bm{u}_{m}(t)-\bm{u}_{m}^{\text{stop}}\right\Vert -\frac{v_{m}^{s}(t)+v_{\max}^{s}}{2}\tau_{0}}{v_{\max}^{s}\tau_{0}}\right\rceil -\left\lceil \frac{\left\Vert \bm{u}_{m}(t+1)-\bm{u}_{m}^{\text{stop}}\right\Vert +\frac{v_{m}^{s}(t+1)-v_{\max}^{s}}{2}\tau_{0}}{v_{\max}^{s}\tau_{0}}\right\rceil -1\nonumber \\
\overset{(a)}{\geq} & \left\lfloor \frac{-\frac{v_{m}^{s}(t)+v_{m}^{s}(t+1)}{2}-\frac{v_{m}^{s}(t)+v_{\max}^{s}}{2}-\frac{v_{m}^{s}(t+1)-v_{\max}^{s}}{2}}{v_{\max}^{s}}\right\rfloor -1\nonumber \\
= & \left\lfloor \frac{-v_{m}^{s}(t)-v_{m}^{s}(t+1)}{v_{\max}^{s}}\right\rfloor -1\nonumber \\
\geq & -3,\label{eq:dif_req_t}
\end{align}
where (a) is due to the fact that $\lceil i\rceil-\lceil j\rceil\geq\lfloor i-j\rfloor$.
Bringing (\ref{eq:dif_req_t}) into (\ref{eq:time_diff_up}), we can
get that $\phi_{m}(t+1)\geq\phi_{m}(t)-4.$ }

\textcolor{black}{By assuming the maximum energy consumption $\overline{E}$
in each time slot, we can obtain that $\psi_{m}(t)-\psi_{m}(t+1)\leq4\overline{E}$
due to $\phi_{m}(t)-\phi_{m}(t+1)\leq4$. Hence, we have $\psi_{m}(t+1)\geq\psi_{m}(t)-4\overline{E}$.}

\textcolor{blue}{}

\bibliographystyle{IEEEtran}
\bibliography{AoI,multi_UAV2}

\begin{thebibliography}{10}
\providecommand{\url}[1]{#1}
\csname url@samestyle\endcsname
\providecommand{\newblock}{\relax}
\providecommand{\bibinfo}[2]{#2}
\providecommand{\BIBentrySTDinterwordspacing}{\spaceskip=0pt\relax}
\providecommand{\BIBentryALTinterwordstretchfactor}{4}
\providecommand{\BIBentryALTinterwordspacing}{\spaceskip=\fontdimen2\font plus
\BIBentryALTinterwordstretchfactor\fontdimen3\font minus
  \fontdimen4\font\relax}
\providecommand{\BIBforeignlanguage}[2]{{%
\expandafter\ifx\csname l@#1\endcsname\relax
\typeout{** WARNING: IEEEtran.bst: No hyphenation pattern has been}%
\typeout{** loaded for the language `#1'. Using the pattern for}%
\typeout{** the default language instead.}%
\else
\language=\csname l@#1\endcsname
\fi
#2}}
\providecommand{\BIBdecl}{\relax}
\BIBdecl

\bibitem{yi2020deep}
M.~Yi, X.~Wang, J.~Liu, Y.~Zhang, and B.~Bai, ``{Deep Reinforcement Learning
  for Fresh Data Collection in UAV-assisted IoT Networks},'' in \emph{Proc.
  {{IEEE INFOCOM WKSHPS}}}, July 2020.

\bibitem{Mozaffari_TutorialUAV_WN_2019}
M.~{Mozaffari}, W.~{Saad}, M.~{Bennis}, Y.~{Nam}, and M.~{Debbah}, ``{A
  Tutorial on UAVs for Wireless Networks: Applications, Challenges, and Open
  Problems},'' \emph{IEEE Commun. Surv. Tutorials}, vol.~21, no.~3, pp.
  2334--2360, 2019.

\bibitem{zongyong_UAV5G_tutorial2019}
Y.~{Zeng}, Q.~{Wu}, and R.~{Zhang}, ``{Accessing From the Sky: A Tutorial on
  UAV Communications for 5G and Beyond},'' \emph{Proc. IEEE}, vol. 107, no.~12,
  pp. 2327--2375, Dec 2019.

\bibitem{JGong_TminUAV_DC_2018}
J.~{Gong}, T.~{Chang}, C.~{Shen}, and X.~{Chen}, ``{Flight Time Minimization of
  UAV for Data Collection Over Wireless Sensor Networks},'' \emph{IEEE J. Sel.
  Areas Commun.}, vol.~36, no.~9, pp. 1942--1954, 2018.

\bibitem{Yzeng_EEUAV_trajectory_optimization_2017}
Y.~{Zeng} and R.~{Zhang}, ``{Energy-Efficient UAV Communication With Trajectory
  Optimization},'' \emph{IEEE Trans. Wireless Commun.}, vol.~16, no.~6, pp.
  3747--3760, 2017.

\bibitem{CHLiu_EE_FairCC_DRL_2018}
C.~H. {Liu}, Z.~{Chen}, J.~{Tang}, J.~{Xu}, and C.~{Piao}, ``{Energy-Efficient
  UAV Control for Effective and Fair Communication Coverage: A Deep
  Reinforcement Learning Approach},'' \emph{IEEE J. Sel. Areas Commun.},
  vol.~36, no.~9, pp. 2059--2070, 2018.

\bibitem{RDing_3DTra-Freq-EE-FairC-DRL_2020TWC}
R.~{Ding}, F.~{Gao}, and X.~S. {Shen}, ``{3D UAV Trajectory Design and
  Frequency Band Allocation for Energy-Efficient and Fair Communication: A Deep
  Reinforcement Learning Approach},'' \emph{IEEE Trans. Wireless Commun.},
  vol.~19, no.~12, pp. 7796--7809, 2020.

\bibitem{PM_EE-VNF-DRL_TGCN2022}
M.~Pourghasemian, M.~R. Abedi, S.~S. Hosseini, N.~Mokari, M.~R. Javan, and
  E.~A. Jorswieck, ``{AI-Based Mobility-Aware Energy Efficient Resource
  Allocation and Trajectory Design for NFV Enabled Aerial Networks},''
  \emph{IEEE Trans. Green Commun. Netw.}, pp. 1--1, 2022.

\bibitem{Abd_elmagid-AoI_role_in_IoT-2019}
M.~A. Abd-Elmagid, N.~Pappas, and H.~S. Dhillon, ``{On the Role of Age of
  Information in the Internet of Things},'' \emph{IEEE Commun. Mag.}, vol.~57,
  no.~12, pp. 72--77, 2019.

\bibitem{BOM_TheRoleofUAV-IoTNetworkinFutureWildfireDetection_2021IoT}
O.~M. Bushnaq, A.~Chaaban, and T.~Y. Al-Naffouri, ``{The Role of UAV-IoT
  Networks in Future Wildfire Detection},'' \emph{IEEE Internet Things J.},
  vol.~8, no.~23, pp. 16\,984--16\,999, 2021.

\bibitem{S.Kaul_Mini_AoI_VehicularNet}
S.~{Kaul}, M.~{Gruteser}, V.~{Rai}, and J.~{Kenney}, ``{Minimizing Age of
  Information in Vehicular Networks},'' in \emph{Proc. IEEE 8th Annu. Commun.
  Soc. Conf. Sensor, Mesh, Ad Hoc Commun. Netw.}, 2011, pp. 350--358.

\bibitem{AM_AoI-VNF-CompoundAC-MADRL-JSAC2021}
M.~Akbari, M.~R. Abedi, R.~Joda, M.~Pourghasemian, N.~Mokari, and
  M.~Erol-Kantarci, ``{Age of Information Aware VNF Scheduling in Industrial
  IoT Using Deep Reinforcement Learning},'' \emph{IEEE J. Sel. Areas Commun.},
  vol.~39, no.~8, pp. 2487--2500, 2021.

\bibitem{Guo:2022ub}
C.~Guo, X.~Wang, L.~Liang, and G.~Y. Li, ``Age of information, latency, and
  reliability in intelligent vehicular networks,'' \emph{IEEE Network}, pp.
  1--8, 2022.

\bibitem{JHu_UAVs_traj_DRL_2020}
J.~{Hu}, H.~{Zhang}, L.~{Song}, R.~{Schober}, and H.~V. {Poor}, ``{Cooperative
  Internet of UAVs: Distributed Trajectory Design by Multi-Agent Deep
  Reinforcement Learning},'' \emph{IEEE Trans. Commun.}, vol.~68, no.~11, pp.
  6807--6821, 2020.

\bibitem{WFY_UAV-device-multiDRL_TCOM_2021}
F.~Wu, H.~Zhang, J.~Wu, Z.~Han, H.~V. Poor, and L.~Song, ``{UAV-to-Device
  Underlay Communications: Age of Information Minimization by Multi-Agent Deep
  Reinforcement Learning},'' \emph{IEEE Trans. Commun.}, vol.~69, no.~7, pp.
  4461--4475, 2021.

\bibitem{liMinimizingPacketExpiration2019}
W.~Li, L.~Wang, and A.~Fei, ``Minimizing {{Packet Expiration Loss With Path
  Planning}} in {{UAV}}-{{Assisted Data Sensing}},'' \emph{IEEE Wirel. Commun.
  Lett.}, vol.~8, no.~6, pp. 1520--1523, Dec. 2019.

\bibitem{JLiu_UAV_AoIWSN}
J.~{Liu}, P.~{Tong}, X.~{Wang}, B.~{Bai}, and H.~{Dai}, ``{UAV-Aided Data
  Collection for Information Freshness in Wireless Sensor Networks},''
  \emph{IEEE Trans. Wireless Commun.}, pp. 1--1, 2020.

\bibitem{Hu_UAV_AoI_chargeSN_IoT_2021}
H.~Hu, K.~Xiong, G.~Qu, Q.~Ni, P.~Fan, and K.~B. Letaief, ``{AoI-Minimal
  Trajectory Planning and Data Collection in UAV-Assisted Wireless Powered IoT
  Networks},'' \emph{IEEE Internet Things J.}, vol.~8, no.~2, pp. 1211--1223,
  2021.

\bibitem{Jia_AoI_UAV_IoT_2019}
Z.~Jia, X.~Qin, Z.~Wang, and B.~Liu, ``{Age-Based Path Planning and Data
  Acquisition in UAV-Assisted IoT Networks},'' in \emph{Proc. IEEE Int. Conf.
  Commun. Workshops (ICC Workshops)}, 2019, pp. 1--6.

\bibitem{abd-elmagidDeepReinforcementLearning2019}
M.~A. {Abd-Elmagid}, A.~Ferdowsi, H.~S. Dhillon, and W.~Saad,
  ``\BIBforeignlanguage{en}{{Deep {{Reinforcement Learning}} for {{Minimizing
  Age}}-of-{{Information}} in {{UAV}}-Assisted {{Networks}}}},'' in
  \emph{\BIBforeignlanguage{en}{Proc. IEEE Global Commun. Conf. (GLOBECOM)}},
  {Puako, HI, USA}, May 2019.

\bibitem{Ferdowsi_tuyouhua-DRL-UAV-AoI_2021}
A.~Ferdowsi, M.~A. Abd-Elmagid, W.~Saad, and H.~S. Dhillon, ``{Neural
  Combinatorial Deep Reinforcement Learning for Age-Optimal Joint Trajectory
  and Scheduling Design in UAV-Assisted Networks},'' \emph{IEEE J. Sel. Areas
  Commun.}, vol.~39, no.~5, pp. 1250--1265, 2021.

\bibitem{abd-elmagidAveragePeakAgeofInformation2019}
M.~A. {Abd-Elmagid} and H.~S. Dhillon, ``Average {{Peak Age}}-of-{{Information
  Minimization}} in {{UAV}}-assisted {{IoT Networks}},'' \emph{IEEE Trans. Veh.
  Technol.}, vol.~68, no.~2, pp. 2003--2008, 2019.

\bibitem{Zhang_AoI_UAVrelay_energy_2020}
S.~Zhang, H.~Zhang, Z.~Han, H.~V. Poor, and L.~Song, ``{Age of Information in a
  Cellular Internet of UAVs: Sensing and Communication Trade-Off Design},''
  \emph{IEEE Trans. Wireless Commun.}, vol.~19, no.~10, pp. 6578--6592, 2020.

\bibitem{SMY_AoIEnergyAwareUAVAssistedDataCollectIoTNet:DRL_IoT2021}
M.~Sun, X.~Xu, X.~Qin, and P.~Zhang, ``{AoI-Energy-Aware UAV-Assisted Data
  Collection for IoT Networks: A Deep Reinforcement Learning Method},''
  \emph{IEEE Internet Things J.}, vol.~8, no.~24, pp. 17\,275--17\,289, 2021.

\bibitem{Samir_AoI_mltiUAV_transport_DDPG_2020}
M.~Samir, C.~Assi, S.~Sharafeddine, D.~Ebrahimi, and A.~Ghrayeb, ``{Age of
  Information Aware Trajectory Planning of UAVs in Intelligent Transportation
  Systems: A Deep Learning Approach},'' \emph{IEEE Trans. Veh. Technol.},
  vol.~69, no.~11, pp. 12\,382--12\,395, 2020.

\bibitem{Abedin_multiUAV_maxEF_AoIconstraint_DQN_2020}
S.~F. Abedin, M.~S. Munir, N.~H. Tran, Z.~Han, and C.~S. Hong, ``{Data
  Freshness and Energy-Efficient UAV Navigation Optimization: A Deep
  Reinforcement Learning Approach},'' \emph{IEEE Trans. Intell. Transp. Syst.},
  pp. 1--13, 2020.

\bibitem{OOS_MultiUAVAoI-WPCN-MultiagentRL_2021INFOCOMWKSH}
O.~S. Oubbati, M.~Atiquzzaman, A.~Lakas, A.~Baz, H.~Alhakami, and W.~Alhakami,
  ``{Multi-UAV-enabled AoI-aware WPCN: A Multi-agent Reinforcement Learning
  Strategy},'' in \emph{Proc. IEEE Int. Conf. Comput. Commun. Workshops
  (INFOCOM WKSHPS)}, 2021, pp. 1--6.

\bibitem{sn-energy-bonuli1_JSAC2016}
A.~Baknina and S.~Ulukus, ``{Optimal and Near-Optimal Online Strategies for
  Energy Harvesting Broadcast Channels},'' \emph{IEEE J. Sel. Areas Commun.},
  vol.~34, no.~12, pp. 3696--3708, 2016.

\bibitem{sn-energy-bonuli2_JSAC2016}
D.~Shaviv and A.~Ozgur, ``{Universally Near Optimal Online Power Control for
  Energy Harvesting Nodes},'' \emph{IEEE J. Sel. Areas Commun.}, vol.~34,
  no.~12, pp. 3620--3631, 2016.

\bibitem{sn-energy-bonuli3_TIT2020}
A.~Arafa, J.~Yang, S.~Ulukus, and H.~V. Poor, ``{Age-Minimal Transmission for
  Energy Harvesting Sensors With Finite Batteries: Online Policies},''
  \emph{IEEE Trans. Inf. Theory}, vol.~66, no.~1, pp. 534--556, 2020.

\bibitem{al2014optimal}
A.~Al-Hourani, S.~Kandeepan, and S.~Lardner, ``{Optimal LAP Altitude for
  Maximum Coverage},'' \emph{IEEE Wireless Commun. Lett.}, vol.~3, no.~6, pp.
  569--572, 2014.

\bibitem{Zhangrui_slot_dis_unchange_TWC2018}
Y.~Zeng, X.~Xu, and R.~Zhang, ``{Trajectory Design for Completion Time
  Minimization in UAV-Enabled Multicasting},'' \emph{IEEE Trans. Wireless
  Commun.}, vol.~17, no.~4, pp. 2233--2246, 2018.

\bibitem{AoI-max_IoT2022}
L.~Liu, K.~Xiong, J.~Cao, Y.~Lu, P.~Fan, and K.~B. Letaief, ``{Average AoI
  Minimization in UAV-Assisted Data Collection With RF Wireless Power Transfer:
  A Deep Reinforcement Learning Scheme},'' \emph{IEEE Internet Things J.},
  vol.~9, no.~7, pp. 5216--5228, 2022.

\bibitem{oliehoek2016concise}
F.~A. Oliehoek, C.~Amato \emph{et~al.}, \emph{{A concise introduction to
  decentralized POMDPs}}.\hskip 1em plus 0.5em minus 0.4em\relax Springer,
  2016, vol.~1.

\bibitem{Q_learning}
C.~J. Watkins and P.~Dayan, ``{Q-learning},'' \emph{Machine learning}, vol.~8,
  no.~3, pp. 279--292, 1992.

\bibitem{sutton2018reinforcement}
R.~S. Sutton and A.~G. Barto, \emph{{Reinforcement learning: An
  introduction}}.\hskip 1em plus 0.5em minus 0.4em\relax MIT press, 2018.

\bibitem{tampuu2017IDQL}
A.~Tampuu, T.~Matiisen, D.~Kodelja, I.~Kuzovkin, K.~Korjus, J.~Aru, J.~Aru, and
  R.~Vicente, ``{Multiagent Cooperation and Competition with Deep Reinforcement
  Learning},'' \emph{PloS one}, vol.~12, no.~4, p. e0172395, 2017.

\bibitem{rashid2018qmix}
T.~Rashid, M.~Samvelyan, C.~S. De~Witt, G.~Farquhar, J.~Foerster, and
  S.~Whiteson, ``{QMIX: Monotonic Value Function Factorisation for Deep
  Multi-Agent Reinforcement Learning},'' \emph{arXiv:1803.11485}, 2018.

\bibitem{chung2014GRU}
J.~Chung, C.~Gulcehre, K.~Cho, and Y.~Bengio, ``{Empirical Evaluation of Gated
  Recurrent Neural Networks on Sequence Modeling},'' \emph{arXiv:1412.3555},
  2014.

\bibitem{SakHa_LSTMcomplexAnalyse_2014}
H.~{Sak}, A.~{Senior}, and F.~{Beaufays}, ``{Long Short-Term Memory Based
  Recurrent Neural Network Architectures for Large Vocabulary Speech
  Recognition},'' \emph{arXiv:1402.1128}, p. arXiv:1402.1128, Feb. 2014.

\bibitem{Wu_QuantizedConNN_2016_CVPR}
J.~Wu, C.~Leng, Y.~Wang, Q.~Hu, and J.~Cheng, ``{Quantized Convolutional Neural
  Networks for Mobile Devices},'' in \emph{Proc. IEEE Conf. Comput. Vis.
  Pattern Recognit. (CVPR)}, June 2016.

\bibitem{Hausknecht2015DRQN}
M.~Hausknecht and P.~Stone, ``{Deep Recurrent Q-Learning for Partially
  Observable MDPs},'' \emph{arXiv:1507.06527}, 2015.

\bibitem{leng2019age}
S.~Leng and A.~Yener, ``{Age of Information Minimization for Wireless Ad Hoc
  Networks: A Deep Reinforcement Learning Approach},'' in \emph{Proc. IEEE
  Global Commun. Conf. (GLOBECOM)}.\hskip 1em plus 0.5em minus 0.4em\relax
  IEEE, 2019, pp. 1--6.

\bibitem{HHM_AoI-traj-plan-data-collec-wireless-power-IoT2021}
H.~Hu, K.~Xiong, G.~Qu, Q.~Ni, P.~Fan, and K.~B. Letaief, ``{AoI-Minimal
  Trajectory Planning and Data Collection in UAV-Assisted Wireless Powered IoT
  Networks},'' \emph{IEEE Internet Things J.}, vol.~8, no.~2, pp. 1211--1223,
  2021.

\bibitem{mozaffari2017mobile}
M.~Mozaffari, W.~Saad, M.~Bennis, and M.~Debbah, ``{Mobile Unmanned Aerial
  Vehicles (UAVs) for Energy-Efficient Internet of Things Communications},''
  \emph{IEEE Trans. Wireless Commun.}, vol.~16, no.~11, pp. 7574--7589, 2017.

\bibitem{foerster2016parameter_sharing}
J.~Foerster, I.~A. Assael, N.~De~Freitas, and S.~Whiteson, ``{Learning to
  Communicate with Deep Multi-Agent Reinforcement Learning},'' in \emph{Proc.
  29th Adv. Neural Inf. Process. Syst. (NeurIPS)}, 2016, pp. 2137--2145.

\end{thebibliography}

\end{document}